\documentclass[aps,prd,nofootinbib,groupedaddress]{revtex4}
\usepackage{graphicx}
\usepackage{epsfig}
\usepackage{dcolumn}
\usepackage{amsfonts}
\usepackage{amssymb}
\usepackage{bm}
\usepackage{amsmath}
\def\Journal#1#2#3#4{{#1} {\bf#2}, #3 (#4)}

\def\NPA{{\rm Nucl. Phys.} A}
\def\NPB{{\rm Nucl. Phys.} B}
\def\PLB{{\rm Phys. Lett.}  B}
\def\PRL{\rm Phys. Rev. Lett.}
\def\PRD{{\rm Phys. Rev.} D}
\def\PRC{{\rm Phys. Rev.} C}

\def\EPJC{{\rm Eur. Phys. J.} C}

\def\ep{\epsilon}

\def\la{\langle}
\def\ra{\rangle}

\def\lam{\lambda}
\def\al{\alpha}

\def\be{\begin{equation}}
\def\ee{\end{equation}}
\def\bea{\begin{eqnarray}}
\def\eea{\end{eqnarray}}
\begin{document}
\title{Consistency of the light-front quark model with chiral symmetry in the pseudoscalar meson analysis}
\author{ Ho-Meoyng Choi\\
{\em Department of Physics, Teachers College, Kyungpook National University,
     Daegu, Korea 702-701}\\
         Chueng-Ryong Ji\\
{\em Department of Physics, North Carolina State University,
Raleigh, NC 27695-8202} }
\begin{abstract}
We discuss the link between the chiral symmetry of QCD and the numerical results of the light-front quark model (LFQM),
analyzing both the two-point and three-point functions of a pseudoscalar meson from the perspective of the vacuum
fluctuation consistent with the chiral symmetry of QCD. The two-point and three-point functions are exemplified
in this work by the twist-2 and twist-3 distribution amplitudes of a pseudoscalar meson and the pion elastic form factor, respectively.
The present analysis of the pseudoscalar meson commensurates with the previous analysis of the vector meson
two-point function and fortifies our observation that the light-front quark model with effective degrees of freedom
represented by the constituent quark and antiquark may provide the view of effective zero-mode cloud around the quark
and antiquark inside the meson. Consequently, the constituents dressed by the zero-mode cloud may be expected
to satisfy the chiral symmetry of QCD. Our results appear consistent with this expectation and effectively indicate that
the constituent quark and antiquark in the LFQM may be considered as the dressed constituents including the
zero-mode quantum fluctuations from the vacuum.
\end{abstract}

\maketitle
\section{Introduction}

Hadronic distribution amplitudes (DAs)  provide essential
information on the QCD interaction of quarks, antiquarks and gluons inside the hadrons and
play an essential role in applying QCD to hard exclusive processes.
They are the longitudinal projection of the hadronic wave functions obtained by integrating the transverse
momenta of the fundamental constituents~\cite{BL80,ER80,CZ84}.
These nonperturbative quantities are defined as vacuum-to-hadron matrix elements
of particular nonlocal quark or  quark-gluon operators and
thus encode important information on bound states in strong interaction physics.
It has motivated many studies using various nonperturbative models
\cite{BF, Ball99, BBL, CJ_DA, PR, NK06, HWZ04, HZW05}
and led to develop distinct phenomenological models over the past two decades.
Among them, the light-front quark model (LFQM) appears to be one of the most efficient and effective tools in hadron physics
as it takes advantage of the distinguished features of the light-front dynamics (LFD)~\cite{BPP}.
In particular, the LFD carries the maximum number~(seven) of the kinetic
(or interaction independent) generators and thus the less effort in dynamics is
necessary in order to get the QCD solutions that reflect the full Poincar$\acute{e}$ symmetries.
Moreover, the rational energy-momentum dispersion relation  of LFD, namely
$p^-=({\bf p}^2_\perp + m^2) / p^+$, yields the sign correlation between the LF energy
$p^-(=p^0-p^3)$ and the LF longitudinal momentum $p^+(=p^0 + p^3)$ and
leads to the suppression of quantum fluctuations of the vacuum, sweeping
the complicated vacuum fluctuations into the zero-modes in the limit of
$p^+ \rightarrow 0$~\cite{Zero1,Zero2,Zero3}.
This simplification is a remarkable advantage in LFD and facilitates the partonic interpretation of the amplitudes.
Based on the advantages of the LFD, the LFQM  has been developed~\cite{CJ_99} and subsequently applied for
various meson phenomenologies such as the mass spectra of both heavy and light mesons~\cite{CJ_Bc},
the decay constants, DAs, form factors and generalized parton distributions~\cite{Jaus90,Jaus99,Jaus03,Cheng04,CJ_99,CJ_DA,Choi07,BPP,MF12,CJ_Bc,CJ_PV}.

Despite these successes in reproducing the general features
of the data, however, it has proved very difficult to obtain direct connection
between the LFQM and QCD.
To discuss the link between the chiral symmetry of QCD and the numerical results of the LFQM,
we recently presented a self-consistent covariant description of vector meson decay constants and chirality-even quark-antiquark
DAs up to twist 3 in LFQM~\cite{CJ_V14}.
Although the meson decay amplitude described by a two-point function could be regarded as one of the simplest possible physical observables,
it is interesting that this apparently simple amplitude bears
abundant fundamental information on QCD vacuum dynamics and chiral symmetry.
In particular, we discussed the zero-mode issue in the LFQM prediction of vector meson decay constants from the perspective of the vacuum fluctuation consistent with the chiral symmetry of QCD and extended the exactly
solvable manifestly covariant Bethe-Salpeter~(BS) model calculation to the more phenomenologically
accessible realistic LFQM.

To discuss the nature of the LF zero-mode in meson decay amplitude,
we may denote the total LF longitudinal momentum of the meson, $P^+ = k_Q^+ + k_{\bar Q}^+$, where
$k_Q^+$ and $k_{\bar Q}^+$ are the individual quark and antiquark LF longitudinal momenta, respectively.
Similarly, the total LF energy $P^-$ is shared by the individual quark and antiquark LF energies $k_Q^-$ and $k_{\bar Q}^-$,
i.e. $P^- = k_Q^- + k_{\bar Q}^-$.
For the LF energy integration of the two-point function over $k_Q^-$ or $k_{\bar Q}^-$ to compute the meson decay amplitude,
one may use the Cauchy's theorem for a contour integration and pick up the LF energy pole, e.g. either $[k_Q^-]_{\rm on}$
(i.e. on-shell value of $k_Q^-$) from the quark propagator or $[k_{\bar Q}^-]_{\rm on}$ from the antiquark propagator.
However, it is crucial to note that the poles move to infinity (or fly away in the complex plane) as the LF longitudinal momentum,
either $k_Q^+$ or $k_{\bar Q}^+$, goes to zero~\cite{BDJM}.
Unless the contribution from the pole flown into infinity vanishes, it must be kept in computing the physical
observable that must reflect the full Poincar$\acute{e}$ symmetries.
Since such contribution, if it exists, appears either from $k_Q^+=0$ and $k_{\bar Q}^+=P^+$ or
from $k_{\bar Q}^+=0$ and $k_Q^+=P^+$, we call it as the zero-mode contribution.
In the case of two-point function for the computation of the meson decay constant,
the zero-mode contribution is thus locked into a single point of the LF longitudinal momentum, i.e. either $k_Q^+=0$ where
$k_{\bar Q}^+=P^+$ or $k_{\bar Q}^+=0$ where $k_Q^+=P^+$. As one of the
constituents of the meson carries the entire momentum $P^+$ of the meson in this case, the other constituent carries the zero LF
longitudinal momentum that can be regarded as the zero-mode quantum fluctuation linked to the vacuum.
This link is due to a pair creation of particles with zero LF longitudinal momenta from the vacuum and
it is important to capture the vacuum effect for the consistency with the chiral symmetry properties of the strong
interactions~\cite{JMT2013}.
With this link, the zero-mode contribution in the meson decay process can be considered effectively
as the effect of vacuum fluctuation consistent with the chiral symmetry of the strong interactions.
In this respect, the LFQM with effective degrees of freedom represented by the constituent quark and antiquark
may be linked to the QCD since the zero-mode link to the QCD vacuum may provide the view of an effective zero-mode cloud
around the quark and antiquark inside the meson.  Although the constituents are dressed by the zero-mode cloud, they are still expected to
satisfy the chiral symmetry consistent with the QCD. Our numerical results~\cite{CJ_V14}
were indeed consistent with this expectation and effectively indicated that
the constituent quark and antiquark in the standard LFQM~\cite{CJ_99,CJ_DA,Choi07,Jaus90,Cheng97,Hwang10,Kon}
could be considered as the dressed constituents including the zero-mode quantum fluctuations from the QCD vacuum.

Since the constituent quark and antiquark used in the LFQM have already absorbed the zero-mode cloud,
the zero-mode contribution in the LFQM may not be as explicit as in the manifestly covariant model calculation although it effectively provides the consistency with the chiral symmetry. The standard light-front (SLF) approach of the LFQM, with which the observables
are directly computed in 3-dimensional LF momentum space, is not amenable to determine the zero-mode contribution by itself and thus it has been a common practice to utilize an exactly solvable manifestly covariant BS model to check the existence (or absence) of the zero-mode as one can pin down the zero mode exactly in the manifestly covariant BS model.
Within the covariant BS model, we indeed found the nonvanishing zero modes in the vector meson decay amplitude and identified the corresponding zero-mode operators that can be applied to the LFQM.
We also found the self-consistent correspondence relations (see e.g. Eq.~(49) in~\cite{CJ_V14}) between the covariant BS model
and the LFQM that allow the substitution of the radial and spin-orbit wave functions
of the exactly solvable model by the more phenomenologically accessible model wave functions
that can be provided by the LFQM analysis of meson masses~\cite{CJ_99}.
What is remarkable in our finding~\cite{CJ_V14} is that
the nonvanishing zero-mode contributions as well as the instantaneous ones to the vector meson decay
amplitude appeared in the covariant BS model now vanish explicitly when the phenomenological
wave function such as the Gaussian wave function in LFQM is used through the aforementioned correspondence relation.
In another words, the decay constants and the quark DAs of vector mesons can be obtained
only from the on-mass-shell valence contribution within the framework of the standard
LFQM~\cite{CJ_99,Kon,Jaus90,Jaus91,Cheng97,CCP,Card95,Hwang10,CJK02,Choi07}
using the Gaussian radial wave function and they still satisfy the chiral symmetry consistent with
the QCD.

One of the key ingredients for this finding is the isospin symmetry, namely,
the symmetric DAs for the equal quark and antiquark bound state mesons (e.g. $\rho$ meson).
Under the exchange of the LF longitudinal momentum fraction of the quark and antiquark, $x \leftrightarrow (1-x)$,
the DA of the meson with the two equal-mass constituents must be symmetric, $\phi(x)=\phi(1-x)$.
We exploited this fundamental constraint anticipated from the isospin symmetry
to identify the correct DAs in LFQM. The twist-2 and twist-3 DAs of the $\rho$ meson
obtained only from the on-mass-shell valence constituents in LFQM~\cite{CJ_V14} not only satisfy
this constraint anticipated from the isospin symmetry but also reproduce the correct
asymptotic DAs in the chiral symmetry limit.
Knowing that the higher-twist DAs may come from the contributions of the higher Fock-states
such as pair terms as well as the transverse motion of constituents
in the leading twist components~\cite{BF,Ball99,BBL}, we should further attest that our LFQM formulation for
the twist-3 DA is indeed simple without involving zero modes and thus the connected contributions
to the current arising from the vacuum disappear in our LFQM calculation, yet preserves
all the necessary constraints anticipated from the isospin symmetry and the chiral symmetry.

The purpose of this work is to extend our previous work to analyze the decay amplitude related with twist-3 DAs of a pseudoscalar meson within the LFQM and show that the analysis of pseudoscalar mesons fortify our previous conclusion drawn from the vector meson case~\cite{CJ_V14}. That is, the treacherous points such as the zero-mode and the instantaneous contributions present in the covariant BS model disappear in the standard LFQM with
the Gaussian radial wave function but nevertheless satisfy the chiral symmetry.
The twist-3 DAs of a pseudoscalar meson appear to play an important role in constraining our LFQM to be consistent with the conclusion drawn from our previous analysis of the vector meson decay constant. The twist-2 DA of a pseudoscalar meson has been analyzed in our previous work of LFQM~\cite{CJ_DA}.
Essentially, there are two independent twist-3 two-particle DAs of a pseudoscalar meson, namely,
$\phi^{\cal P}_{3;M}$ and $\phi^\sigma_{3;M}$~\cite{BF,Ball99,BBL,PR,NK06,HWZ04,HZW05}
corresponding to pseudoscalar and tensor channels
of a meson ($M$), respectively.
In this work, we shall study $\phi^{\cal P}_{3;M}$ together
with the twist-2 DA $\phi^{\cal A}_{2;M}$ corresponding to the axial-vector channel for the sake of completeness.

The $\phi^{\cal A}_{2;M}$ and $\phi^{\cal P}_{3;M}$ are defined in terms of the
following matrix elements of gauge invariant nonlocal operators at light-like
separation~\cite{BF,Ball99,BBL}:
\be\label{Deq:1}
\la 0|{\bar q}(z)[z,-z]\gamma^\mu\gamma_5 q(-z)|M(P)\ra
= if_M P^\mu \int^1_0 dx e^{i\zeta P\cdot z} \phi^{\cal A}_{2;M}(x),
\ee
and
\be\label{Deq:2}
\la 0|{\bar q}(z)[z,-z]i\gamma_5 q(-z)|M(P)\ra
= f_M \mu_M \int^1_0 dx e^{i\zeta P\cdot z} \phi^{\cal P}_{3;M}(x),
\ee
where $z^2=0$ and the path-ordered gauge link (Wilson line) $[z,-z]$  for the gluon fields between
the points $-z$ and $z$ is equal to unity in the light-cone gauge $A(z)\cdot z=0$ which we take
throughout our calculation. $P$ is the four-momentum of the meson ($P^2=m^2_M$) and the integration variable $x$ corresponds to the longitudinal momentum fraction
carried by the quark and $\zeta =2x -1$ for the short-hand notation.
The normalization parameter $\mu_M = m^2_M /(m_q + m_{\bar q})$ in Eq.~(\ref{Deq:2})
results from quark condensate. For the pion, $\mu_\pi = -2\la {\bar q}q\ra / f^2_\pi$ from the Gell-Mann-Oakes-Renner relation~\cite{GOR}.
The normalization of the two DAs $\Phi=\{ \phi^{\cal A}_{2;M}, \phi^{\cal P}_{3;M} \}$ is given by
\be\label{Deq:3}
\int^1_0 dx \; \Phi(x) = 1.
\ee
In order to check the existence (or absence) of the zero mode,
we again utilize the same manifestly covariant model used in the analysis of the vector meson
decay constant~\cite{CJ_V14} and then substitute the
vertex function with the more phenomenologically accessible Gaussian radial wave function
provided by our LFQM. We shall show that the analysis of the decay constants and twist-2 and twist-3 two-particle DAs of pseudoscalar mesons confirms our previous conclusion drawn for the vector meson
analysis~\cite{CJ_V14}. Namely, the treacherous points such as the zero-mode and the instantaneous
contributions appeared in the covariant BS model do not show up explicitly in the standard LFQM with
the Gaussian radial wave function but nevertheless satisfy the chiral symmetry.

In addition, we show that our findings of the zero-mode complication in two-point function is
directly applicable to the three-point function with the analysis of the pion elastic form factor.
The analyses of the pion form factor using the plus component ($J^+_{\rm em}$) of the LF currents
have been done in many earlier works~\cite{BCJ01,CJK,CJ06,CJ08} and proved that the pion form
factor is immune to the zero-mode contribution when the plus component of the currents
is used. Particularly, in our LFQM analysis of the pion form factor~\cite{CJ06,CJ08}, we have
shown that the usual power-law behavior of the pion form factor obtained in the perturbative
QCD analysis can also be attained by taking negligible quark masses in our nonperturbative LFQM
analysis, confirming the anti-de Sitter space geometry/conformal field theory (AdS/CFT)
correspondence~\cite{BT04}.
In this work, we analyze the pion form factor using the perpendicular components
($J^\perp_{\rm em}$) of the currents. Within the covariant BS model, we find that the form
factor obtained in the $q^+=0$ frame with $J^\perp_{\rm em}$ receives only the valence
contribution including both the on-mass-shell quark propagating part and the off-mass-shell
instantaneous part without involving a zero mode. Applying this to the LFQM, we find
that the nonvanishing instantaneous contribution appeared in the BS model does not appear and
just the on-mass-shell propagating part contributes
in the LFQM. This example of the three-point function provides an evidence that the
conclusion drawn in the LFQM analysis of the two-point function is also applicable to
the three-point function.

The paper is organized as follows. In Sec.~\ref{sec:II}, we discuss the decay amplitude
of a pseudoscalar meson described by the two-point function and the pion form factor described
by three-point function in an exactly solvable model based on the covariant BS model of
(3+1)-dimensional fermion field theory. We mainly perform our LF calculation for
the decay amplitude corresponding to the twist-3 DA $\phi^{\cal P}_{3;M}$ and
the pion form factor using $J^\perp_{\rm em}$  and check the LF
covariance of them within the covariant BS model. Especially, we discuss how to identify the zero-mode contribution
and find the corresponding zero-mode operator.
In Sec.~\ref{sec:III}, we present the standard LFQM with the gaussian wave function and discuss the correspondence linking the manifestly covariant model to the standard LFQM.
The self-consistent covariant descriptions of the meson decay constants as well as the
twist-2 and twist-3 two-particle DAs of pseudoscalar mesons in the standard LFQM are given in this section.
In Sec.~\ref{sec:IV}, we present our numerical results for the explicit demonstration
of our findings. Summary and discussion follow in Sec.~\ref{sec:V}.

\section{Manifestly Covariant Model}
\subsection{Two-point function: decay amplitude}
\label{sec:II}
\begin{figure}
\begin{center}
\includegraphics[height=3.5cm, width=7cm]{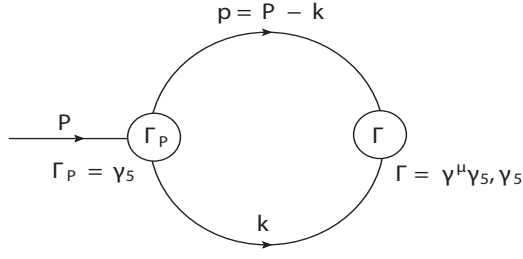}
\caption{\label{fig1} Feynman diagram for the one-quark-loop evaluation of the meson
decay amplitude in the momentum space. }
\end{center}
\end{figure}
Defining the local matrix elements
${\cal M}_\al \equiv \la 0|{\bar q}\Gamma_\al q|M(P)\ra (\al={\cal A, P})$
for axial-vector ($\Gamma_{\cal A} =\gamma^\mu\gamma_5$) and
pseudoscalar ($\Gamma_{\cal P} =i\gamma_5$) channels of Eqs.~(\ref{Deq:1})~and~(\ref{Deq:2}),
we write the one-loop approximation (see Fig.~\ref{fig1}) as a momentum integral
\be\label{Deq:4}
{\cal M}_\al = N_c
\int\frac{d^4k}{(2\pi)^4} \frac{H_0} {N_p N_k} S_\al,
\ee
where $N_c$ denotes the number of colors.
The denominators $N_p (= p^2 -m^2_q +i\varepsilon)$
and $N_k(= k^2 - m^2_{\bar q}+i\varepsilon)$ come from the quark propagators
of mass $m_q$ and $m_{\bar q}$ carrying the internal four-momenta $p =P -k$ and $k$, respectively.
In order to regularize the covariant loop,
we use the usual multipole ansatz~\cite{Jaus99,CJ_V14,MF97,SS} for the $q{\bar q}$ bound-state vertex function
$H_0=H_0(p^2,k^2)$ of a meson:
\be\label{Deq:5}
H_0(p^2,k^2) = \frac{g}{N_\Lambda^n},
\ee
where $N_\Lambda  = p^2 - \Lambda^2 +i\varepsilon$, and $g$ and $\Lambda$ are constant parameters.
Although the vertex function $H_0$ could be symmetrized in the four momenta of the
constituent quarks for further study, we take a simplest possible regularization in this work
as a tool to analyze the zero-mode complication in the exactly solvable model.
In the same vein, although the power $n$ for the multipole ansatz could be $n\geq 2$ to regularize the loop integral,
we take the lowest possible power $n=2$ since our qualitative results in terms of the zero-mode issue
do not depend on the value of $n$.

The trace term $S_\alpha$  in Eq.~(\ref{Deq:4}) is given by
\be\label{Deq:6}
 S_\al  =  {\rm Tr}\left[\Gamma_\al\left(\slash \!\!\!p+m_q \right)
 \gamma_5
 \left(-\slash \!\!\!k + m_{\bar q} \right) \right].
\ee
We have already computed the matrix element ${\cal M_A}$ of the axial-vector
channel in the Appendix B of Ref.~\cite{CJ_V14} and have shown that ${\cal M_A}$ (i.e. the decay constant
of a pseudoscalar meson) obtained from the plus component of the currents is immune to the zero mode. Therefore, we shall discuss the pseudoscalar channel and the associated
twist-3 DA $\phi^{\cal P}_{3;M}$ in this work.
After a  little manipulation, we obtain the manifestly covariant result for
${\cal M_P}$ as follows
\bea\label{Deq:7}
{\cal M}_{\cal P}^{\rm cov} &=& \frac{N_c g}{4\pi^2} \int^1_0 dx\int^{1-x}_0 dy (1-x-y)
\biggl\{ \frac{y(1-y)m^2_M + m_q m_{\bar q}}{C^2_{\rm cov}}
- \frac{2}{C_{\rm cov}} \biggr\},
\eea
where
$C_{\rm cov} = y(1-y) m^2_M - x m^2_q - y m^2_{\bar q} - (1-x-y) \Lambda^2$.

For the LF calculation in parallel with the manifestly covariant one,
we separate the trace term $S_{\cal P}$ in Eq.~(\ref{Deq:6})
into the on-mass-shell propagating part $[S_{\cal P}]_{\rm on}$ and the off-mass-shell instantaneous part $[S_{\cal P}]_{\rm inst}$ via $\slash\!\!\!q
=\slash\!\!\!q_{\rm on} + \frac{1}{2}\gamma^+(q^- - q^-_{\rm on})$
as
\be\label{Deq:8A}
S_{\cal P} = [S_{\cal P}]_{\rm on} + [S_{\cal P}]_{\rm inst},
\ee
where
$[S_{\cal P}]_{\rm on} = 4 ( p_{\rm on}\cdot k_{\rm on} + m_q m_{\bar q} )$
and
$[S_{\cal P}]_{\rm inst} = 2( p^+ \Delta_k^- +  k^+ \Delta_p^- )$
with $\Delta_{q}^-=q^- -q_{\rm on}^-$. We note that the metric convention
$a\cdot b =\frac{1}{2}(a^+b^- + a^-b^+)-{\bf a}_\perp\cdot{\bf b}_{\perp}$ is used
in our analysis.
Furthermore, we take the reference frame where ${\bf P}_\perp =0$, i.e.,
$P=( P^+, M^2/P^+, 0)$. In this case, the LF energies of the on-mass-shell
quark and antiquark are given by
$ p^-_{\rm on} = ({\bf k}^2_\perp + m^2_q)/ xP^+$ and
$ k^-_{\rm on} = ({\bf k}^2_\perp + m^2_{\bar q})/(1-x) P^+$, respectively,
where $x=p^+/P^+$ is the LF longitudinal momentum fraction of the quark.

For the integration over $k^-$ in Eq.~(\ref{Deq:4}), one may close the contour in the lower
half of the complex $k^-$ plane and pick up the residue at $k^-=k^-_{\rm on}$
in the region $0< k^+ < P^+$ (or $0 < x < 1$).
We denote the valence contribution to ${\cal M_P}$ which is obtained by taking $k^-=k^-_{\rm on}$
in the region of $0<x<1$ as  $[{\cal M_P}]^{\rm LF}_{\rm val}$ that is
given by
\be\label{Deq:10}
 [{\cal M_P}]^{\rm LF}_{\rm val} = \frac{N_c}{16\pi^3}\int^{1}_0
 \frac{dx}{(1-x)} \int d^2{\bf k}_\perp
 \chi(x,{\bf k}_\perp) [S_{\cal P}]_{\rm val},
\ee
where
\be\label{Deq:11}
\chi(x,{\bf k}_\perp) = \frac{g}{[x (m_M^2 -M^2_0)][x (m_M^2 - M^2_{\Lambda})]^n},
\ee
with $n=2$ and
\bea
\label{Deq:12}
 M^2_{0(\Lambda)} &=& \frac{ {\bf k}^{2}_\perp + m^2_q(\Lambda^2)}{x}
 + \frac{ {\bf k}^{2}_\perp + m^2_{\bar q}}{1-x}.
\eea
Here, the trace term for the valence contribution, i.e.
$[S_{\cal P}]_{\rm val}=[S_{\cal P}]_{\rm on} + 2 k^+ \Delta_p^-$, is given by
\be\label{Deq:13}
[S_{\cal P}]_{\rm val}=2 [ M^2_0 - (m_q - m_{\bar q})^2 + (1-x)E_{\rm E.B.} ],
\ee
where the binding energy term $E_{\rm E.B.}=m^2_M - M^2_0$ stems from the instantaneous contribution.
We find numerically that $[{\cal M_P}]^{\rm LF}_{\rm val}$ in Eq.~(\ref{Deq:10}) is not identical to the
manifestly covariant result ${\cal M_P}^{\rm cov}$ in Eq.~(\ref{Deq:7}). This indicates that the
decay amplitude ${\cal M_P}$ receives the LF zero-mode contribution.
The LF zero-mode contribution to ${\cal M_P}$ comes from the singular
$p^-$ (or equivalently $1/x$) term in $S_{\cal P}$ in the limit
of $x\to 0$ when $p^-=p^-_{\rm on}$, i.e.
\be \label{Deq:14}
\lim_{x\to 0}S_{\cal P}(p^-=p^-_{\rm on})= 2p^-.
\ee
The necessary prescription to identify zero-mode operator corresponding to
$p^-$ is analogous to that derived in the previous analyses of weak transition form factor
calculations~\cite{Jaus99,CJ_Bc,CJ_PV}, except that there is no momentum transfer $q$ dependence.
As extensively discussed in the previous works~\cite{Jaus99,CJ_Bc,CJ_PV,CJ_V14},
we now identify the zero-mode operator $[S_{\cal P}]_{\rm Z.M.}$
by replacing $p^-$ with $-Z_2$ in Eq.~(\ref{Deq:14}), i.e.
\be\label{Deq:15}
[S_{\cal P}]_{\rm Z.M.} = 2 (-Z_2),
\ee
where
$Z_2 = x E_{\rm E.B.} + m^2_q - m^2_{\bar q} + (1-2x)m_M^2$.
 This zero-mode operator $[S_{\cal P}]_{\rm Z.M.}$ can be effectively included
in the valence region as follows
\bea\label{Deq:16}
 [{\cal M_P}]^{\rm LF}_{\rm full}
 &=& \frac{N_c}{16 \pi^3}\int^{1}_0
 \frac{dx}{(1-x)} \int d^2{\bf k}_\perp
 \chi(x,{\bf k}_\perp)[S_{\cal P}]_{\rm full},
\eea
where
$[S_{\cal P}]_{\rm full}= [S_{\cal P}]_{\rm val} + [S_{\cal P}]_{\rm Z.M.}$ and it is given by
\be\label{Deq:17}
[S_{\cal P}]_{\rm full} = 4 [ xM^2_0 + m_q (m_{\bar q}-m_q) ].
\ee
It can be checked that
Eq.~(\ref{Deq:16}) is identical to the manifestly covariant result of
Eq.~(\ref{Deq:7}).

Although the amplitude ${\cal M_A} = if_M P^\mu$ for the axial vector channel
is proven to be immune to the zero mode when the plus component ($\mu=+$) of the currents is used
and its form is given in Ref.~\cite{CJ_V14}, we display it here again for completeness in the form of a pseudoscalar meson
decay constant:
\be\label{Deq:18}
f^{\rm LF}_M = \frac{N_c}{4\pi^3}\int^1_0\frac{dx}{(1-x)}\int d^2{\bf k}_\perp
\chi(x,{\bf k}_\perp) {\cal A},
\ee
where ${\cal A}= (1-x) m_q + x m_{\bar q}$.

\subsection{Three-point function: Pion electromagnetic form factor}
The electromagnetic form factor of a pion is defined by the matrix elements of the
current $J^\mu_{\rm em}$:

\be\label{Eeq:1}
\la P'|J^\mu_{\rm em}|P\ra = e_m(P+P')^\mu F_\pi(q^2),
\ee
where $e_m$ is the charge of the meson and $q^2=(P-P')^2$ is the square of the four momentum transfer.

\begin{figure}
\begin{center}
\includegraphics[height=3.5cm, width=15cm]{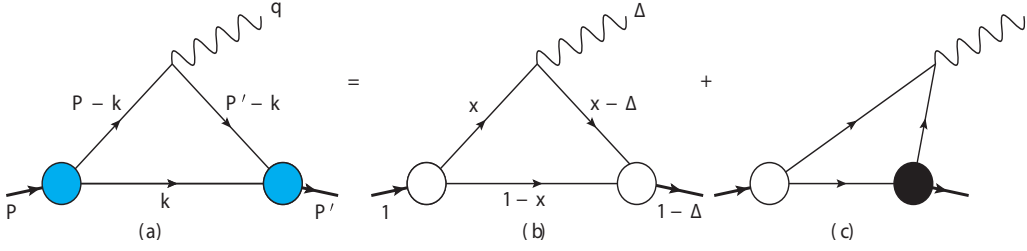}
\caption{\label{fig2} The covariant triangle diagram (a) corresponds to the
sum of the LF valence diagram (b) and the nonvalence diagram (c). The large white
and black blobs at the meson-quark vertices in (b) and (c) represent the ordinary
LF wave function and the nonvalence wave function vertices, respectively. }
\end{center}
\end{figure}
The covariant diagram shown in Fig.~\ref{fig2}(a) to describe the pion form factor
is in general equivalent to the sum of the LF valence diagram [Fig.~\ref{fig2}(b)]
and the nonvalence diagram [Fig.~\ref{fig2}(c)]. The matrix element
${\cal J}^\mu\equiv\la P'|J^\mu_{\rm em}|P\ra$ obtained from
the covariant diagram of Fig.~\ref{fig2}(a) is given by
 \be\label{Eeq:2}
 {\cal J}^\mu = iN_c \int\frac{d^4k}{(2\pi)^4}
 \frac{H'_0 H''_0}{N_{p_1} N_{k} N_{p_2}}S^\mu,
 \ee
where
 \be\label{Eeq:3}
 S^\mu = {\rm Tr}[\gamma_5(\not\!p_1 + m)\gamma^\mu
 (\not\!p_2 + m)\gamma_5(-\not\!k + m)]
 \ee
with $p_1=P-k$ and $p_2=P'-k$. Here, we take $m_q=m_{\bar q}=m$ for the pion.
The vertex functions are given by $H'_0=H'_0(p^2_1,k^2)=g/(N'_\Lambda)^n$ and
$H''_0=H''_0(p^2_2,k^2)=g/(N''_\Lambda)^n$ with
$N'_\Lambda (N''_\Lambda) = p_1^2 (p_2^2) -\Lambda^2+i\ep$. In this case, we take
the power $n$ for the multipole ansatz to be simply 1, since our qualitative results in conjunction
with the zero-mode issue do not depend on the value of $n$.
The rest of the denominator factor $N_{p}$ from the intermediate quark
propagator with momentum $p=(p_1,p_2,k)$ is given by $N_{p}= p^2 - m^2 + i\ep$.

Using the usual Feynman parametrization,
we obtain the manifestly covariant result as follows
\bea\label{Eeq:5}
  F^{\rm cov}_\pi(q^2) &=& \frac{N_c g^2}{8\pi^2 (\Lambda^2-m^2)^2}\int^1_0 dx\int^{1-x}_0 dy
\biggl\{ [ 3(x+y) -4]\ln\biggl(\frac{C_{\Lambda m}C_{m \Lambda}}
{C_{m m} C_{\Lambda \Lambda}} \biggr)
\nonumber\\
&&
 + [ (1-x-y)^2 (x + y)m_\pi^2 + x y(2 - x -y) q^2 - (x + y)m^2 + 2m^2 ] C
 \biggr\},
\eea
where
$C=(1/C_{\Lambda \Lambda} -1/C_{\Lambda m} - 1/C_{m \Lambda} + 1/C_{m m})$ and
$C_{ab} = (1-x-y)(x + y)m_\pi^2 + xy q^2- (x a^2 + y b^2) - (1-x-y)m^2$.

Since the LFD analysis has already shown~\cite{BCJ01,MFPS02} that the pion form factor is immune
to the zero-mode contribution when the
plus component of the currents is used, we shall now explore the perpendicular components ($\mu=\perp$)
of the currents to see if the treacherous points such as the zero-modes exist or not.
In order to check the existence/absence of
the zero-mode contribution to the hadronic matrix element given by Eq.~(\ref{Eeq:2}),
we first choose $q^+ >0$ frame and then take $q^+\to 0$ limit.
In the $q^+>0$ frame, the covariant diagram Fig.~\ref{fig2}(a) corresponds the sum of the LF
valence diagram Fig.~\ref{fig2}(b) defined in $0<k^+<P'^+$ region and the nonvalence diagram Fig.~\ref{fig2}(c) defined in $P'^+ < k^+ <P^+$ region. The large white
and black blobs at the meson-quark vertices in (b) and (c) represent the ordinary
LF wave function and the nonvalence wave function vertices~\cite{BCJ01,BCJ03}, respectively.
Defining $\Delta =q^+/P^+$ and the longitudinal momentum fraction factor
$x=p^+_1/P^+$ ($1-x=k^+/P^+$) for the struck (spectator) quark,
we should note that the nonvalence region (i.e. $0<x<\Delta$) of integration shrinks
to the end point $x=0$ in the $q^+\to 0$ (i.e. $\Delta\to 0$) limit.
The virtue of taking $q^+=0$ frame is to obtain the form factor by calculating only the
valence diagram (i.e. $0 < x <1$) because the nonvalence diagram does not contribute
if the integrand is free from the singularity in $p^-_1\sim 1/x$.
However, if the integrand has a singularity as $x\to 0$,
then one should also take into account this nonvanishing contribution that
we call zero-mode contribution.

In the $q^+=0$ frame with ${\bf P}_\perp =0$, the photon momentum is transverse to the
direction of the incident pion with the spacelike  momentum transfer
${\bf q}^2_\perp\equiv Q^2=-q^2$.
In this frame, one obtains the relations between the current matrix
elements and the pion form factor as follows:
 \bea\label{Eeq:7}
&&   F^{\rm LF}_\pi(Q^2) = \frac{ {\cal J}^+}{ 2P^+}({\rm for}\;\mu=+),
\nonumber\\
&& F^{\rm LF}_\pi(Q^2) =
-\frac{ {\cal J}^\perp \cdot {\bf q}_\perp}{ {\bf q}^2_\perp}({\rm for}\; \mu=\perp).
 \eea
The trace term in Eq.~(\ref{Eeq:3}) can again be separated into on-mass-shell propagating part
and off-mass-shell instantaneous one as $S^\mu = S^\mu_{\rm on} + S^\mu_{\rm inst}$, where
\bea\label{Eeq:8}
  S^\mu_{\rm on} &=&
4 [
 p^\mu_{1\rm on} (p_{2\rm on}\cdot k_{\rm on} + m^2)
 - k^\mu_{\rm on} (p_{1\rm on}\cdot p_{2\rm on} - m^2)
+ p^\mu_{2\rm on} (p_{1\rm on}\cdot k_{\rm on} + m^2)
  ],
 \eea
 and
  \be\label{Eeq:9}
  S^\mu_{\rm inst} =
      2 \Delta^-_{p_1} ( p^\mu_{2\rm on} k^+_{\rm on} - p^+_{2\rm on} k^\mu_{\rm on})
    + 2 \Delta^-_{p_2} ( p^\mu_{1\rm on} k^+_{\rm on} - p^+_{1\rm on} k^\mu_{\rm on} )
    + 2 \Delta^-_{k}  ( p^\mu_{1\rm on} p^+_{2\rm on} + p^+_{1\rm on} p^\mu_{2\rm on} ).
 \ee
Note that Eq.~(\ref{Eeq:9}) is valid only for $\mu=+$ or $\perp$.

In the valence region $0<k^+<P'^+$ (or $0 < x < 1$) of $q^+\to 0$ limit,
the pole $k^-=k^-_{\rm on}$ is located in the lower half of the complex $k^-$-plane.
Performing the LF energy $k^-$ integration of Eq.~(\ref{Eeq:2}),
we obtain the valence contribution to ${\cal J}^\mu$ as
  \bea\label{Eeq:10}
[{\cal J}^\mu]_{\rm val}^{\rm LF} &=&
\frac{N_c}{16\pi^3}\int^1_0 \frac{dx}{(1-x)}\int d^2{\bf k}_\perp
\chi(x,{\bf k}_\perp) \chi' (x, {\bf k}^\prime_\perp)
 S^\mu_{\rm val} ,
  \eea
where $S^\mu_{\rm val}=S^\mu_{\rm on} +
S^\mu_{\rm inst}(\Delta^-_k=0)$  and
${\bf k}^\prime_\perp = {\bf k}_\perp + (1-x){\bf q}_\perp$.
The LF vertex function $\chi$ of the initial state is given by Eq.~(\ref{Deq:11}) but with
$n=1$~\footnote{In this form factor analysis, it is sufficient to consider only the case of
a monopole form of the vertex function ($n=1$), since our qualitative results do not depend on
the value of $n$. }.
The final state vertex function $\chi'$ is equal to $\chi(x, {\bf k}_\perp\to {\bf k'}_\perp)$.

From Eqs.~(\ref{Eeq:7})-(\ref{Eeq:9}), we get the LF valence contributions to the pion form factor
\be\label{Eeq:10a}
 [F_\pi]^{\rm LF(+)}_{\rm val}(Q^2)= \frac{N_c}{8\pi^3}\int^1_0 \frac{dx}{(1-x)^2}
 \int d^2{\bf k}_\perp \chi_1 (x, {\bf k}_\perp)
  \chi_2(x, {\bf k}^\prime_\perp)
   ( {\bf k}_\perp\cdot{\bf k}^\prime_\perp + m^2 ),
 \ee
for $\mu=+$ and
 \be\label{Eeq:10b}
[F_\pi]^{\rm LF(\perp)}_{\rm val}(Q^2) = \frac{N_c}{8\pi^3}\int^1_0 \frac{dx}{(1-x)}
 \int d^2{\bf k}_\perp \chi (x, {\bf k}_\perp)
 \chi'(x, {\bf k}^{\prime}_\perp)
  \biggl[
    (1-x) m_M^2 + x M^2_0
   + \frac{ {\bf k}_\perp\cdot{\bf q}_\perp }{{\bf q}^2_\perp} ( 2 m_M^2 + {\bf q}^2_\perp)
  \biggr],
  \ee
for $\mu=\perp$, respectively. We note that while
the valence contribution for the plus current comes solely from the on-shell
propagating part (i.e. $[F_\pi]^{\rm LF(+)}_{\rm val}=[F_\pi]^{\rm LF(+)}_{\rm on}$),
the valence contribution for the perpendicular currents results not only from the on-shell propagating part
but also from the instantaneous part (i.e.
$[F_\pi]^{\rm LF(\perp)}_{\rm val}=[F_\pi]^{\rm LF(\perp)}_{\rm on}
+ [F_\pi]^{\rm LF(\perp)}_{\rm inst}$), where
\be\label{Eeq:11a}
[F_\pi]^{\rm LF(\perp)}_{\rm on}(Q^2) = \frac{N_c}{8\pi^3}\int^1_0 \frac{dx}{(1-x)}
 \int d^2{\bf k}_\perp \chi (x, {\bf k}_\perp)
 \chi'(x, {\bf k}^\prime_\perp)
 \frac{( {\bf k}_\perp\cdot{\bf k}^\prime_\perp + m^2 )}{x(1-x)}
  \biggl[ 1 + 2\frac{ {\bf k}_\perp\cdot{\bf q}_\perp }{{\bf q}^2_\perp}
  \biggr ],
  \ee
and
 \be\label{Eeq:11b}
[F_\pi]^{\rm LF(\perp)}_{\rm inst}(Q^2) = \frac{N_c}{8\pi^3}\int^1_0 \frac{dx}{(1-x)}
 \int d^2{\bf k}_\perp \chi(x, {\bf k}_\perp)
 \chi'(x, {\bf k}^\prime_\perp)
  \biggl[ \frac{ {\bf k}^\prime_\perp\cdot{\bf q}_\perp }{{\bf q}^2_\perp}(m_M^2-M^2_0)
  + \frac{ {\bf k}_\perp\cdot{\bf q}_\perp }{{\bf q}^2_\perp}(m_M^2-M^{\prime 2}_0)
  \biggr ].
  \ee
We should note that while both
$[F_\pi]^{\rm LF(\perp)}_{\rm on}$ and $[F_\pi]^{\rm LF(\perp)}_{\rm inst}$
are infinite, $[F_\pi]^{\rm LF(\perp)}_{\rm val}$ is finite due to the cancellation of the infinity.
Furthermore, we find from our numerical computation that the three results
$[F_\pi]^{\rm LF(+)}_{\rm val}$ in Eq.~(\ref{Eeq:10a}),
$[F_\pi]^{\rm LF(\perp)}_{\rm val}$ in Eq.~(\ref{Eeq:10b}) and
the manifestly covariant result $F^{\rm cov}_\pi$ in Eq.~(\ref{Eeq:5}) are
identical with each other. That is, in this exactly solvable model,
the pion form factor obtained from either
the plus component ($J^+_{\rm em}$) of the currents or
the perpendicular components ($J^\perp_{\rm em}$) of the currents
is immune to the zero-mode contribution.

\section{Application to Standard Light-Front Quark Model}
\label{sec:III}
In the standard LFQM~\cite{CJ_99,Kon,Jaus90,Jaus91,Cheng97,CCP,Card95,Hwang10,CJK02,Choi07}, the
wave function of a ground state pseudoscalar meson ($J^{\rm PC}=0^{-+}$)
as a $q\bar{q}$ bound state is given by
\be\label{QM1}
\Psi_{\lam{\bar\lam}}(x,{\bf k}_{\perp})
={\phi_R(x,{\bf k}_{\perp})\cal R}_{\lam{\bar\lam}}(x,{\bf k}_{\perp}),
\ee
where $\phi_R$ is the radial wave function and the
spin-orbit wave function ${\cal R}_{\lam{\bar\lam}}$
with the helicity $\lam({\bar\lam})$ of a quark(antiquark)
is obtained by the interaction-independent Melosh transformation~\cite{Melosh}
from the ordinary spin-orbit wave function assigned by the quantum numbers $J^{PC}$.

We use the Gaussian wave function for $\phi_R$, which is given by
\be\label{QM2}
\phi_R(x,{\bf k}_{\perp})=
\frac{4\pi^{3/4}}{\beta^{3/2}} \sqrt{\frac{\partial
k_z}{\partial x}} {\rm exp}(-{\vec k}^2/2\beta^2),
\ee
where $\vec{k}^2={\bf k}^2_\perp + k^2_z$ and $\beta$ is the variational parameter
fixed by the analysis of meson mass spectra~\cite{CJ_99}.
The longitudinal component $k_z$ is defined by $k_z=(x-1/2)M_0 +
(m^2_{\bar q}-m^2_q)/2M_0$, and the Jacobian of the variable transformation
$\{x,{\bf k}_\perp\}\to {\vec k}=({\bf k}_\perp, k_z)$ is given by
\be\label{QM3}
\frac{\partial k_z}{\partial x}
= \frac{M_0}{4 x (1-x)} \biggl\{ 1-
\biggl[\frac{m^2_q - m^2_{\bar q}}{M^2_0}\biggr]^2\biggr\}.
\ee
The covariant form of the spin-orbit wave function ${\cal R}_{\lam{\bar\lam}}$
is given by
\be\label{QM4}
{\cal R}_{\lam{\bar\lam}}
=\frac{\bar{u}_{\lam}(p_q)\gamma_5 v_{{\bar\lam}}( p_{\bar q})}
{\sqrt{2}[M^{2}_{0}-(m_q -m_{\bar q})^{2}]^{1/2}},
\ee
and it satisfies
$\sum_{\lam{\bar\lam}}{\cal R}_{\lam{\bar\lam}}^{\dagger}{\cal R}_{\lam{\bar\lam}}=1$.
Thus, the normalization of our wave function is then given by
\be\label{QM6}
1=\sum_{\lam{\bar\lam}}\int\frac{dx d^2{\bf k}_\perp}{16\pi^3}
|\Psi_{\lam{\bar\lam}}(x,{\bf k}_{\perp})|^2
=
\int\frac{dx d^2{\bf k}_\perp}{16\pi^3}
|\phi_R(x,{\bf k}_{\perp})|^2.
\ee

In our previous analysis of the decay constant and the twist-2 and twist-3 DAs of
a vector meson~\cite{CJ_V14},
we have shown that standard light-front (SLF) results of the LFQM is obtained by the
the replacement of the LF vertex function $\chi$ in the BS model with the Gaussian wave function
$\phi_R$ as follows~(see Eq.~(49)~in~\cite{CJ_V14}):
\be\label{Eeq:12}
 \sqrt{2N_c} \frac{ \chi(x,{\bf k}_\perp) } {1-x}
 \to \frac{\phi_R (x,{\bf k}_\perp) }
 {\sqrt{{\bf k}^2_\perp + {\cal A}^2}}, \; m_M \to M_0,
 \ee
where $m_M\to M_0$ implies that the physical mass $m_M$ included in the integrand of BS
amplitude has to be replaced with the invariant mass $M_0$ since the SLF results of the LFQM
are obtained from the requirement of all constituents being on their respective mass-shell.
The correspondence in Eq.~(\ref{Eeq:12}) is valid again in this analysis of a pseudoscalar meson.
For the final state LF vertex function, one should replace ${\bf k}_\perp$ with ${\bf k'}_\perp$
in Eq.~(\ref{Eeq:12}).

We first apply the correspondence given by Eq.~(\ref{Eeq:12}) to the zero-mode free observables
$f^{\rm LF}_M$ [Eq.~(\ref{Deq:18})] and $[F_\pi]^{\rm LF(+)}_{\rm val}(Q^2)$ [Eq.~(\ref{Eeq:10a})].
Then, we obtain the corresponding SLF results $f^{\rm SLF}_M$  and
$F^{\rm SLF(+)}(Q^2)$ as follows:
\be\label{QM7}
f^{\rm SLF}_{M} = \frac{\sqrt{2N_c}}{{8\pi^3}}\int^1_0 dx \int d^2{\bf k}_\perp
\frac{\phi_R(x,{\bf k}_\perp)}{\sqrt{{\bf k}^2_\perp + {\cal A}^2}}
{\cal A},
\ee
and
\be\label{QM8}
F^{\rm SLF (+)}_\pi(Q^2) = \int^1_0 dx \int\frac{d^2{\bf k}_\perp}{16\pi^3}
\phi_R(x,{\bf k}_\perp)\phi'_R(x,{\bf k}^{\prime}_\perp)
\frac{{\bf k}_\perp\cdot {\bf k}^{\prime}_\perp + m^2}
{\sqrt{{\bf k}^2_\perp + m^2} \sqrt{{\bf k}^{\prime 2}_\perp + m^2}},
\ee
where $\phi_R(\phi'_R)$ is the initial (final) state radial wave function.
Eqs.~(\ref{QM7}) and~(\ref{QM8}) are exactly the same as those previously obtained from the SLF approach,
e.g. see Ref.~\cite{CJ_99,CJ_DA}. We should note that both Eqs.~(\ref{QM7}) and~(\ref{QM8}) are the results
obtained only from the on-mass-shell quark propagators. From Eq.~(\ref{QM7}), we obtain
the twist-2 DA $\phi^{\cal A}_{2;M}(x)$ of a pseudoscalar meson as follows
\be\label{QM9}
\phi^{\cal A}_{2;M}(x) = \frac{\sqrt{2N_c}}{{f^{\rm SLF}_{M}8\pi^3}} \int d^2{\bf k}_\perp
\frac{\phi_R(x,{\bf k}_\perp)}{\sqrt{{\bf k}^2_\perp + {\cal A}^2}}
{\cal A}.
\ee
We now apply the correspondence given by Eq.~(\ref{Eeq:12}) to the decay amplitude
${\cal M_P} (=f_M\mu_M)$ for pseudoscalar channel given by Eq.~(\ref{Deq:16}) to obtain
the corresponding LFQM amplitude:
\bea\label{QM10}
 [{\cal M_P}]^{\rm SLF}_{\rm full}
 &=& \frac{\sqrt{2N_c}}{2\cdot 16 \pi^3}\int^{1}_0 dx
 \int d^2{\bf k}_\perp
 \frac{\phi_R(x,{\bf k}_\perp)}{\sqrt{{\bf k}^2_\perp + {\cal A}^2}}
 [S_{\cal P}]_{\rm full},
\eea
where $[S_{\cal P}]_{\rm full} = 4 [ xM^2_0 + m_q (m_{\bar q}-m_q) ]$.

Interestingly enough, we also found that the result $[{\cal M_P}]^{\rm SLF}_{\rm on}$
obtained only from the on-mass-shell quark propagator $[S_{\cal P}]_{\rm on}=2[M^2_0 -(m_q - m_{\bar q})^2]$
is exactly the same as the full result in Eq.~(\ref{QM10}). This equality
$[{\cal M_P}]^{\rm SLF}_{\rm full}=[{\cal M_P}]^{\rm SLF}_{\rm on}$
can be easily seen from the fact that only the even term in $S_P$ with respect to $x$
survives in the SU(2)symmetry limit~($m=m_q=m_{\bar q}$) since the Gaussian wave function $\phi_R$
and other prefactor ${\sqrt{{\bf k}^2_\perp + m^2}}$ are even in $x$. That is, decomposing
the trace term $[S_{\cal P}]_{\rm full} = 4 x M^2_0 = [2 + 2(2x-1)]M^2_0$ in SU(2) symmetry
limit, one can find that the nonvanishing contribution from $[S_{\cal P}]_{\rm full}$ is exactly the
same as $[S_{\cal P}]_{\rm on}=2M^2_0$.
Knowing that the matrix element ${\cal M}_P$ is related with the twist-3 DA $\phi^{\cal P}_{3;M}(x)$,
the above finding in the SU(2) symmetry limit plays the role of the constraint in obtaining the
correct $\phi^{\cal P}_{3;M}(x)$, i.e. only the solution obtained from $[{\cal M_P}]^{\rm SLF}_{\rm on}$
gives the correct $\phi^{\cal P}_{3;M}(x)$ in our LFQM:
\bea\label{QM11}
 \phi^{\cal P}_{3;M}(x)
 &=& \frac{\sqrt{2N_c}}{f^{\rm SLF}_M\mu_M \cdot 16 \pi^3}
 \int d^2{\bf k}_\perp
 \frac{\phi_R(x,{\bf k}_\perp)}{\sqrt{{\bf k}^2_\perp + {\cal A}^2}}
 [M^2_0 -(m_q - m_{\bar q})^2].
\eea
For the pion $(m=m_q=m_{\bar q})$ case, we should note $\mu_\pi = -2\la {\bar q}q\ra / f^2_\pi$.

Applying the correspondence relation in Eq.~(\ref{Eeq:12}) to the pion form factors
$[F_\pi]^{\rm LF(\perp)}_{\rm val}$ [Eq.~(\ref{Eeq:10b})],
$[F_\pi]^{\rm LF(\perp)}_{\rm on}$~[Eq.~(\ref{Eeq:11a})],
and
$[F_\pi]^{\rm LF(\perp)}_{\rm inst}$[Eq.~(\ref{Eeq:11b})] to obtain the corresponding
form factors $[F_\pi]^{\rm SLF(\perp)}_{\rm val}$,
$[F_\pi]^{\rm SLF(\perp)}_{\rm on}$, and $[F_\pi]^{\rm SLF(\perp)}_{\rm inst}$
in our LFQM,
we find that $[F_\pi]^{\rm SLF(\perp)}_{\rm inst}=0$
and $[F_\pi]^{\rm SLF(\perp)}_{\rm on}=F^{\rm SLF (+)}_\pi$.
The explicit form of $[F_\pi]^{\rm SLF(\perp)}_{\rm on}$ is given
by~\footnote{The equivalence between $[F_\pi]^{\rm SLF(\perp)}_{\rm on}$ and $F^{\rm SLF (+)}_\pi$
can be even checked analytically by changing the transverse momentum variables into symmetric ones
in the integrand as follows: ${\bf k}_\perp = {\bf l}_\perp - (1-x){\bf q}_\perp/2$ and
${\bf k'}_\perp = {\bf l}_\perp + (1-x){\bf q}_\perp/2$.}
\be\label{apLF:19}
[F_\pi]^{\rm SLF(\perp)}_{\rm on} = \int^1_0 \frac{dx}{x} \int\frac{d^2{\bf k}_\perp}{16\pi^3}
 \phi_1 (x, {\bf k}_\perp)\phi_2(x, {\bf k}^\prime_\perp)
 \frac{( {\bf k}_\perp\cdot{\bf k}^\prime_\perp + m^2 )}
 {\sqrt{{\bf k}^2_\perp + m^2} \sqrt{{\bf k}^{\prime 2}_\perp + m^2}}
  \biggl[ 1 + 2\frac{ {\bf k}_\perp\cdot{\bf q}_\perp }{{\bf q}^2_\perp}
  \biggr ].
  \ee

\section{Numerical Results}
\label{sec:IV}

\begin{table}[t]
\caption{The constituent quark mass $m_q$ (in GeV) and the gaussian parameters
$\beta_{q{\bar q}}$ (in GeV) for the linear and HO confining potentials
obtained from the variational principle in our LFQM~\cite{CJ_99,CJ_DA,Choi07}. $q=u$ and $d$.}
\label{t1}
\begin{tabular}{ccccc} \hline\hline
Model & $m_q$ & $m_s$ & $\beta_{q{\bar q}}$ & $\beta_{q{\bar s}}$ \\
\hline
Linear & 0.22~ & 0.45~ & 0.3659~ & 0.3886~  \\
HO & 0.25 &  0.48 & 0.3194 & 0.3419  \\
\hline\hline
\end{tabular}
\end{table}
In our numerical calculations within the standard LFQM, we use two sets of model parameters
(i.e. constituent quark masses $m_q$ and the gaussian parameters $\beta_{q{\bar q}}$) for the
linear and harmonic oscillator~(HO) confining potentials given in Table I, which was obtained from the calculation of meson mass spectra using the variational principle in our LFQM~\cite{CJ_99,CJ_DA,Choi07}.

Our LFQM predictions for the decay constants of $\pi$ and $K$ mesons,
$f^{\rm SLF}_\pi=130 ~[131]$ MeV and $f^{\rm SLF}_K= 161~ [155]$ MeV obtained from the
linear~[HO] potential parameters, are in good agreement with the experimental
data~\cite{PDG}; $f^{\rm Exp.}_\pi=(130.41\pm 0.03 \pm 0.20)$ MeV and
$f^{\rm Exp.}_K =(156.2 \pm 0.3 \pm 0.6 \pm 0.3)$ MeV.
We then obtain the quark condensate $\la q{\bar q}\ra$, which enters the normalization
of twist-3 pion DA $\phi^{\cal P}_{3;\pi}(x)$ given by Eq.~(\ref{QM11}), as
$-(285.8 {\rm MeV})^3 ~[-(263.7 {\rm MeV})^3]$ for the linear~[HO] potential parameters.
Our LFQM results, especially the one obtained from HO parameters, are quite comparable with the commonly
used phenomenological value $\la {\bar q}q\ra = -(250 {\rm MeV})^3$.

Defining the LF wave function $\psi^{\cal A(P)}_{2(3);\pi}(x,{\bf k}_\perp)$
for the twist-2 axial-vector (twist-3 pseudoscalar) channel as
\be\label{N1}
\phi^{\cal A(P)}_{2(3);M}(x)
= \int^\infty_0 d^2{\bf k}_\perp \;\psi^{\cal A(P)}_{2(3);M}(x,{\bf k}_\perp),
\ee
the $n$-th transverse moment is obtained by
\be\label{N2}
\la {\bf k}^n_\perp \ra^{\cal A(P)}_M = \int^\infty_0 d^2{\bf k}_\perp
\int^1_0 dx \;\psi^{\cal A(P)}_{2(3);M}(x,{\bf k}_\perp) {\bf k}^n_\perp.
\ee
The authors in Refs.~\cite{SR_C1, SR_C2} have shown
that the second transverse moment can be given in terms of
the quark condensate $\la \bar q q\ra$ and the mixed
quark-gluon condensates $\la ig\bar{q}\sigma\cdot G q\ra$ :
\be\label{N3}
\la {\bf k}^2_\perp \ra^{\cal A}_\pi =
\frac{5}{36}\frac{\la ig\bar{q}\sigma\cdot G q\ra}{\la \bar q q\ra},\;\;
\la {\bf k}^2_\perp \ra^{\cal P}_\pi =
\frac{1}{4}\frac{\la ig\bar{q}\sigma\cdot G q\ra}{\la \bar q q\ra},
\ee
where $G^a_{\mu\nu}$ is a gluon field strength and $\sigma\cdot G =\sigma_{\mu\nu}G^{\mu\nu}$.
Note that the formula for the axial-vector channel is an approximate
one since the soft pion theorems apply strictly only for the pseudoscalar channel~\cite{PR}.

For the pion case, our results of the second transverse moments for the axial-vector
and the pseudoscalar channels obtained from the linear~[HO] parameters
are $\la {\bf k}^2_\perp \ra^{\cal A}_\pi = (413~{\rm MeV})^2~ [(371~{\rm MeV})^2]$ and
$\la {\bf k}^2_\perp \ra^{\cal P}_\pi = (553~{\rm MeV})^2~ [480~{\rm MeV})^2]$, respectively.
Especially, the ratio
$\la {\bf k}^2_\perp \ra^{\cal A}_\pi / \la {\bf k}^2_\perp \ra^{\cal P}_\pi = 0.558$
obtained from the linear parameters  is in good agreement with the QCD sum-rule result,
5/9~\cite{SR_C1}, and the nonlocal chiral model result, $0.54\sim 0.56$~\cite{PR}.
Using Eq.~(\ref{N3}) for the
pseudoscalar channel, we also estimate the value of the mixed condensate of dimension 5 as
$\la ig\bar{q}\sigma\cdot G q\ra=-(491.1~{\rm MeV})^5~[-(442.2~{\rm MeV})^5]$ for the
linear~[HO] parameters. Especially, the result obtained from the linear
parameters is in an excellent agreement with the the result obtained from the
direct calculation in the instanton model~\cite{PW} which gives
$\la ig\bar{q}\sigma\cdot G q\ra=-(490~{\rm MeV})^5$.
For the kaon case, we obtain
$\la {\bf k}^2_\perp \ra^{\cal A}_K = (457~{\rm MeV})^2~ [412~{\rm MeV})^2] $,
$\la {\bf k}^2_\perp \ra^{\cal P}_K = (582~{\rm MeV})^2~ [510~{\rm MeV})^2] $ and
$\la {\bf k}^2_\perp \ra^{\cal A}_K / \la {\bf k}^2_\perp \ra^{\cal P}_K = 0.617~[0.653]$ for
the linear~[HO] parameters.

\begin{figure}
\begin{center}
\includegraphics[height=7cm, width=7cm]{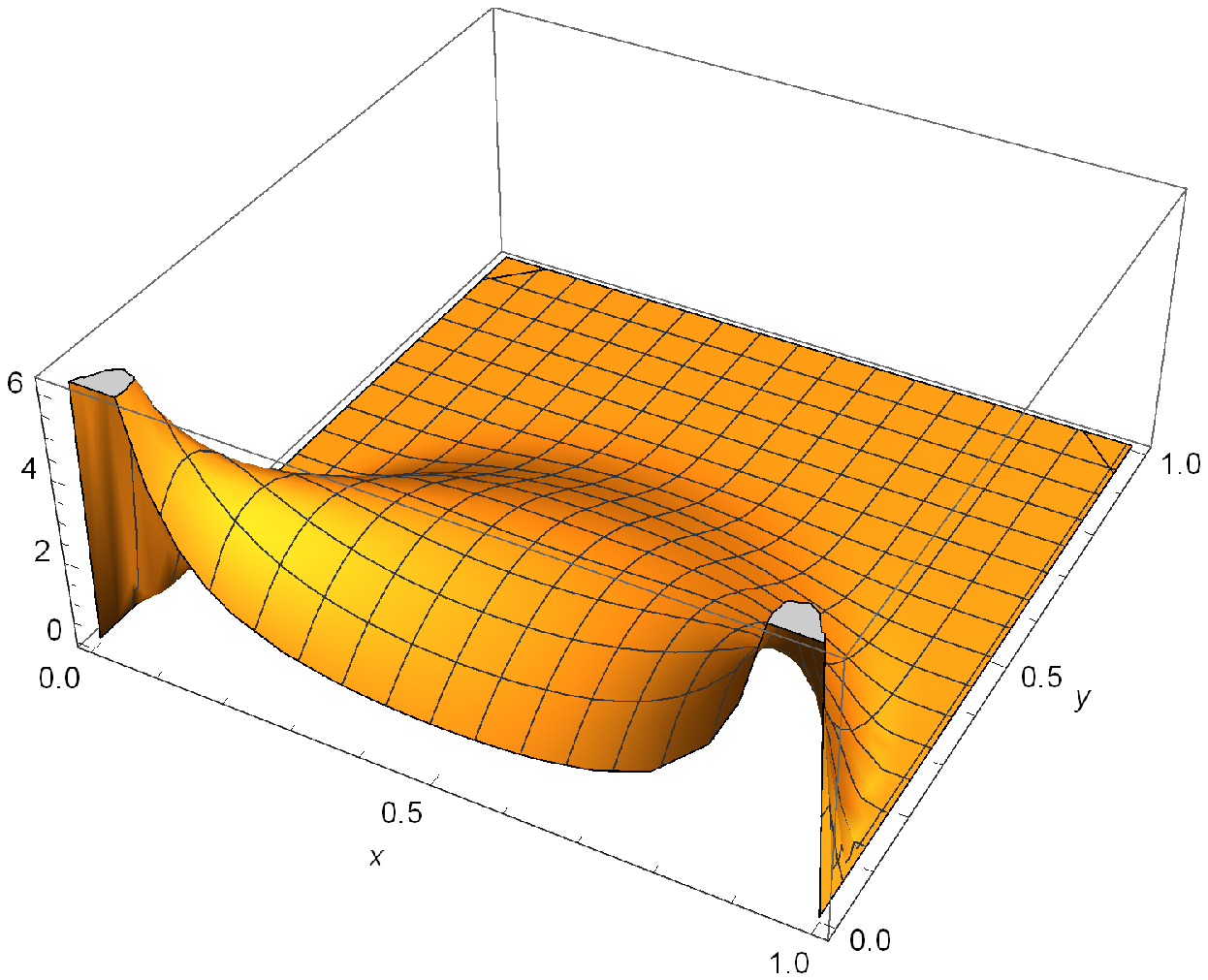}
\hspace{0.5cm}
\includegraphics[height=7cm, width=7cm]{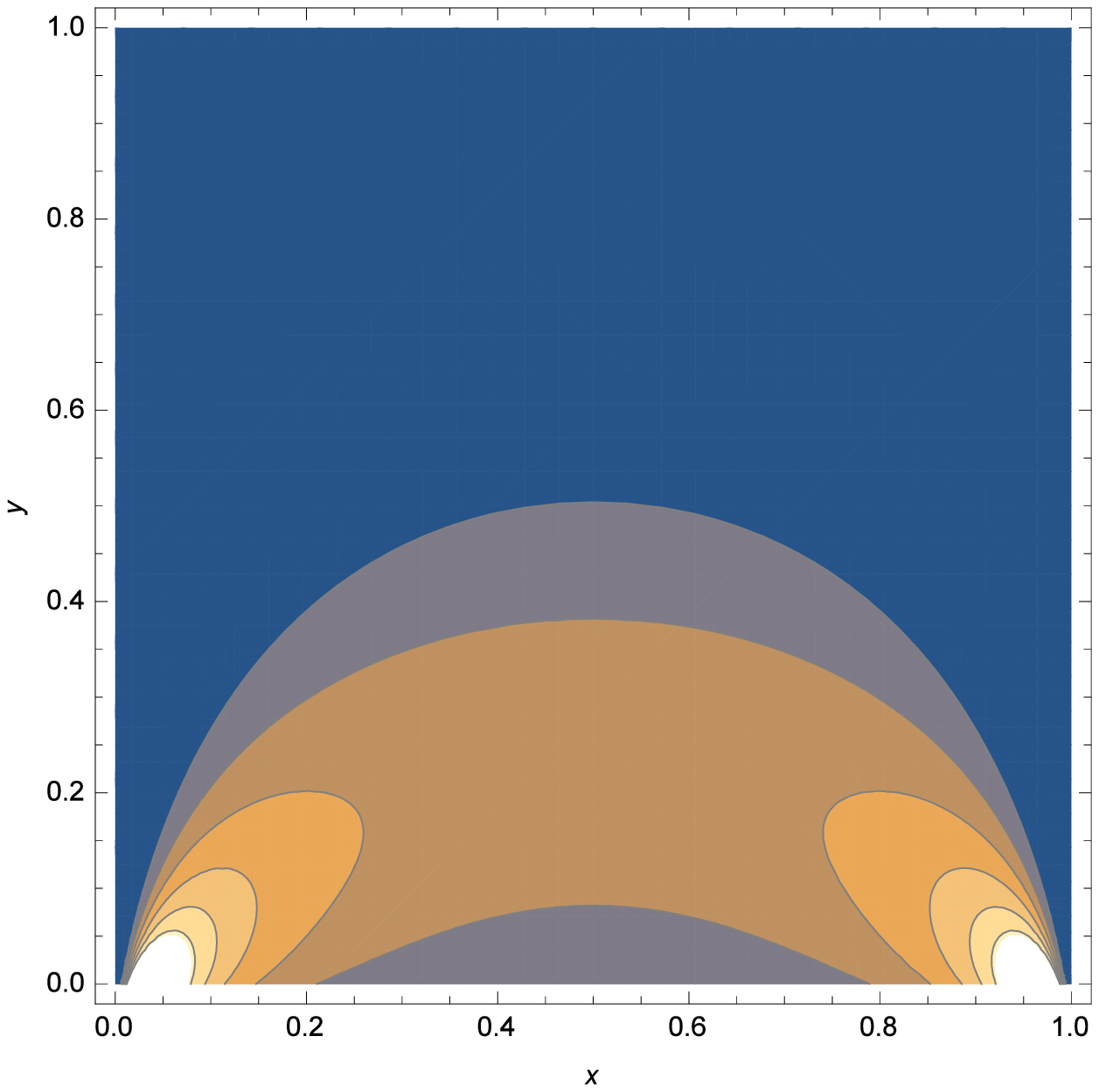}
\\
\includegraphics[height=7cm, width=7cm]{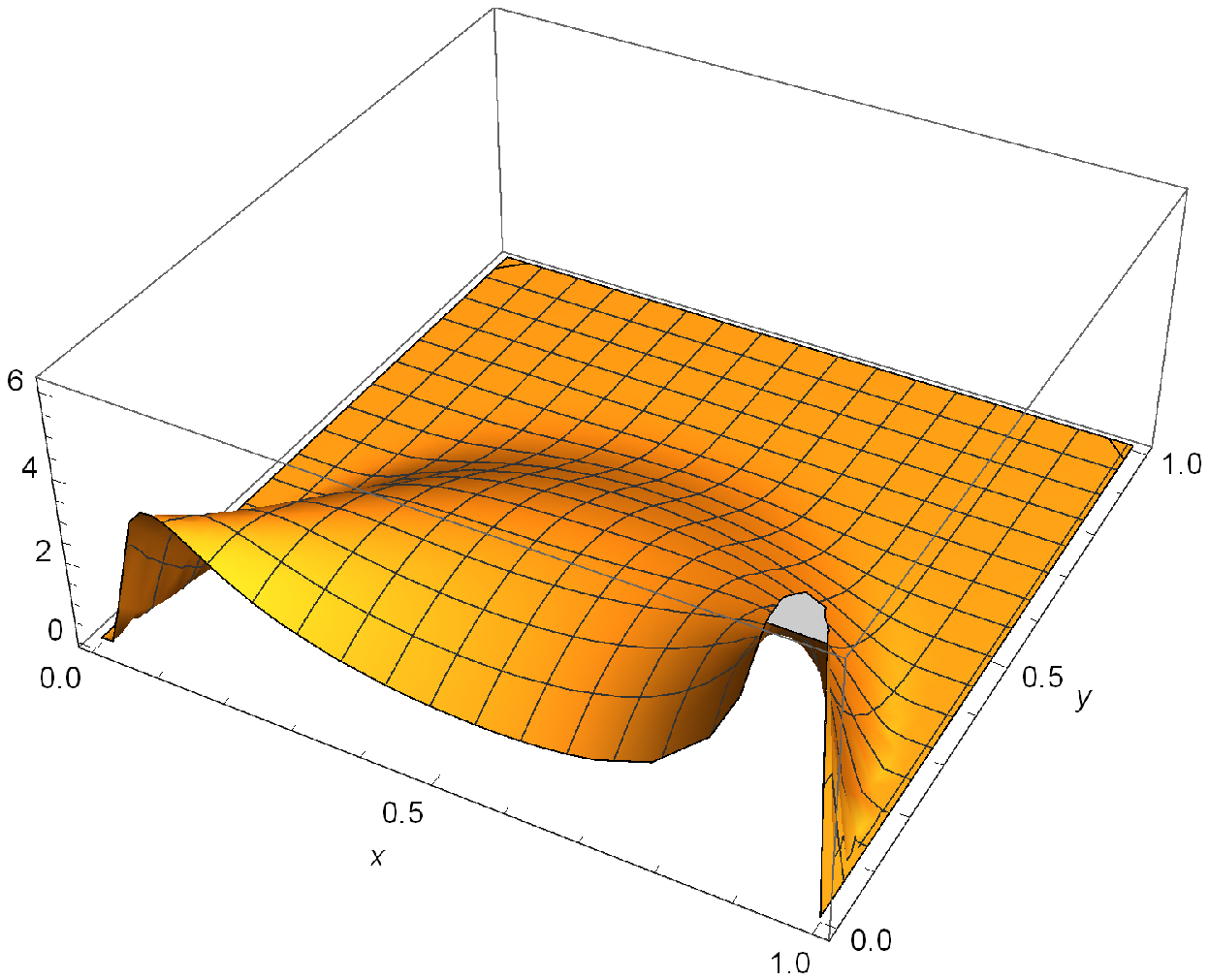}
\hspace{0.5cm}
\includegraphics[height=7cm, width=7cm]{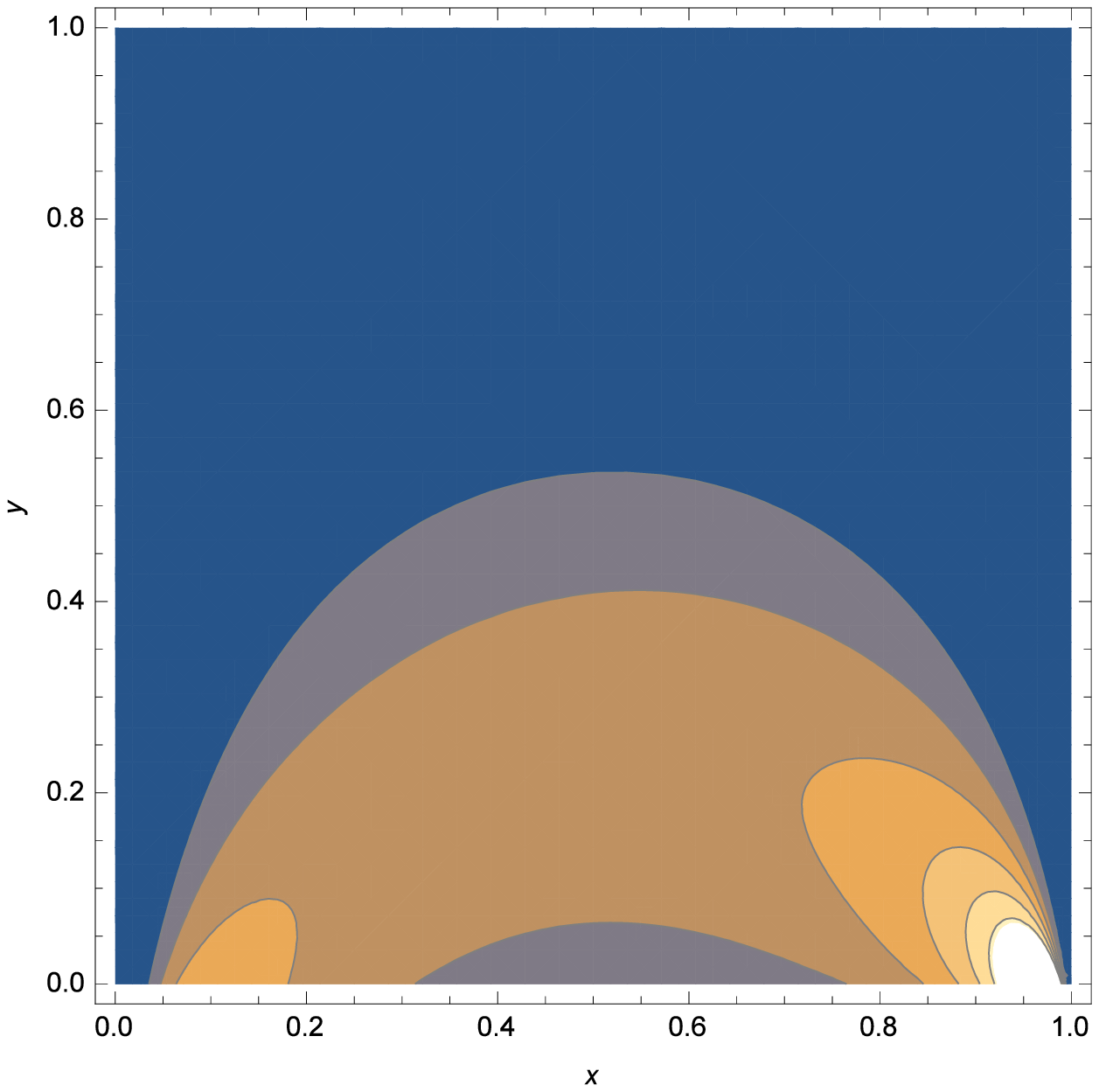}
\caption{\label{fig3} The 3D plot (left panel)
and the corresponding 2D contour plot (right panel )for $\psi^{\cal P}_{3;\pi}(x,y)$ (upper panel)
and $\psi^{\cal P}_{3;K}(x,y)$ (lower panel) obtained
from the linear parameters. In 2D contour plot, darker the regions are, smaller the wave functions are.}
\end{center}
\end{figure}
From the point of view of QCD, the quark DAs of a hadron depend on the scale $\mu$ which
separates nonperturbative and perturbative regimes. In our LFQM, we can associate $\mu$ with
the transverse integration cutoff via $|{\bf k}_\perp|\leq \mu$, which is the usual
way how the normalization scale is defined for the LF wave function~(see, e.g. Ref.~\cite{BL80}).
In order to estimate this
cutoff value, we made a 3-dimensional plot for the LF wave function
$\psi^{\cal P}_{3;\pi(K)}(x,{\bf k}_\perp)$
in the form of $\psi^{\cal P}_{3;\pi(K)}(x,y)$ by changing the variable ${\bf k}^2_\perp = y / (1-y)$
so that $y$ ranges from 0 to 1. Fig.~\ref{fig3} shows the 3D plot (left panel)
and the corresponding 2D contour plot (right panel) for
$\psi^{\cal P}_{3;\pi}(x,y)$ (upper panel)
and $\psi^{\cal P}_{3;K}(x,y)$ (lower panel)
that we obtain with the linear parameters listed in Table~\ref{t1}.
We note that we assign the momentum fraction $x$ for $s$-quark
and $(1-x)$ for the light $u(d)$-quark for $K$ meson case.
In fact, we obtain the twist-3 quark DAs by performing the transverse integration up to infinity
(or equivalently $y$ up to 1) without loss of accuracy due the presence of
Gaussian damping factor. However, as one can see from the contour plots in Fig.~\ref{fig3},
only the range of $0\leq y\leq 0.47$ contribute to the integral for both $\pi$ and $K$ meson cases.
This implies that our cutoff scale corresponds to $y\simeq 0.47$ or equivalently
$\mu\simeq |{\bf k}_\perp|\simeq 1$ GeV for the calculation of the twist-3 $\pi$ and $K$ meson DAs.
Since the twist-2 quark DAs for $\pi$ and $K$ mesons were given in our previous work~\cite{CJ_DA},
we do not show them in this work again but note that the scale $\mu$ for the twist-2 DA is slightly
smaller than that for the twist-3 DA.
Considering both twist-2 and twist-3 DAs of $\pi$
and $K$ mesons, our numerical results show the range of scale $\mu$ as $0.75 \leq \mu\leq 1$ GeV.

\begin{figure}
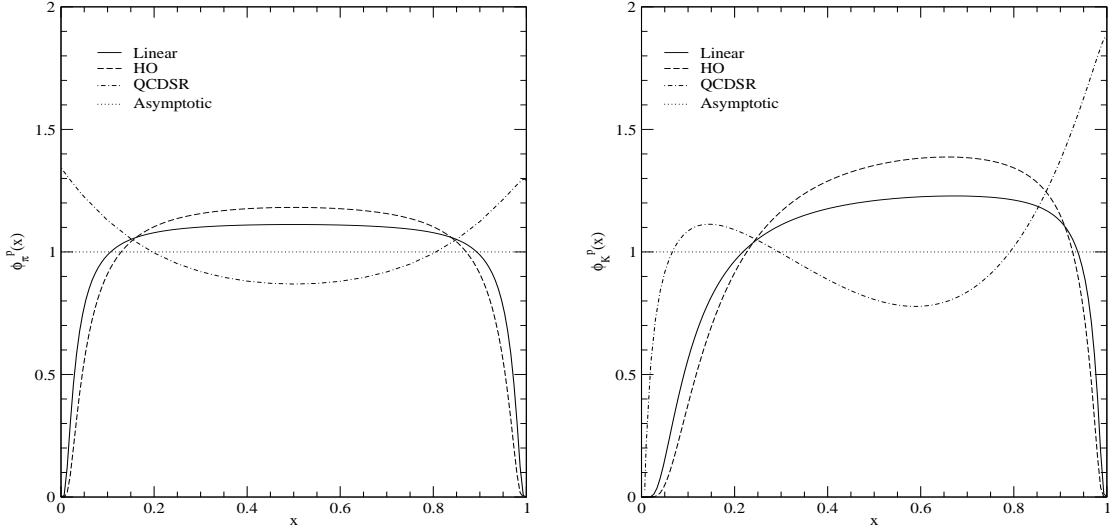

\vspace{0.5cm}
\begin{center}
\includegraphics[height=7cm, width=7cm]{fig4a.eps}
\hspace{0.5cm}
\includegraphics[height=7cm, width=7cm]{fig4b.eps}
\caption{\label{fig4} The twist-3 DAs $\phi^{\cal P}_{3;M}(x)$ for $\pi$ (left panel) and
$K$ (right panel) mesons obtained from the linear~(solid line) and HO (dashed line) parameters
compared with the QCD sum rule result (dot-dashed line)~\cite{BBL} as well as the asymptotic one
(dotted line)~\cite{BF}.}
\end{center}
\end{figure}

We show in Fig.~\ref{fig4} the twist-3 DAs $\phi^{\cal P}_{3;M}(x)$ [see Eq.~(\ref{QM11})]
for $\pi$ (left panel) and
$K$ (right panel) mesons obtained from the linear~(solid line) and HO (dashed line) parameters.
We should note that our LFQM results $\phi^{\cal P}_{3;M}(x)$ are free from the explicit instantaneous as well as zero-mode contributions. The corresponding twist-2 DAs $\phi^{\cal A}_{2;M}$ for $(\pi, K)$ mesons
obtained from our LFQM can be found in~\cite{CJ_DA}.
We also compare our results with the the asymptotic DA $[\phi^{\cal P}_{3;M}]_{\rm as}(x)=1$
(dotted line)~\cite{BF} as well as the QCD sum-rule (QCDSR) prediction (dot-dashed line)~\cite{BBL},
which were obtained at the renormalization scale $\mu =1$ GeV.
For the pion case, our results obtained from both model parameters
not only show the symmetric forms anticipated from the
isospin symmetry but also reproduce the exact asymptotic result
$[\phi^{\cal P}_{3;\pi}]_{\rm as}(x)=1$ in the chiral symmetry ($m_q\to 0$) limit.
This exact asymptotic result $\phi^{\cal P}_{3;\pi}(x)\to [\phi^{\cal P}_{3;\pi}]_{\rm as}(x)$
in the chiral symmetry limit is consistent with the conclusion drawn from our previous
analysis~\cite{CJ_V14} of the twist-2 ($\phi^{||}_{2;\rho}(x)$)
and twist-3 ($\phi^{\perp}_{3;\rho}(x)$) $\rho$ meson DAs.
Remarkably, both DAs reproduce the exact asymptotic DAs in the chiral symmetry limit.
This example shows again that our LFQM prediction satisfies the chiral symmetry consistent
with the QCD as one correctly implements the zero-mode link to the QCD vacuum.
It is also interesting to note that while our results of $\phi^{\cal P}_{3;\pi}(x)$ become zero at the
end points of $x$ unless the asymptotic limit is taken, while the QCD sum-rule result of Ref.~\cite{BBL} does not vanish at the end points.
The main reason for the discrepancy between the two models is that
the QCD sum-rule results are  based on the chiral symmetry($m=0$) limit but our results (unless asymptotic) are based
on the nonvanishing constituent quark model. In the asymptotic limit,
our results exhibit also the nonvanishing behavior at the end points of $x$.
For the $K$ meson case, $\phi^{\cal P}_{3;K}(x)$ obtained from both model parameters are asymmetric
due to the flavor SU(3) symmetry breaking effect and the peak points
located to the right of $x=0.5$ indicate that the $s$-quark carries more longitudinal
momentum fraction than the light $u(d)$-quark.

The twist-2 and twist-3 quark DAs are usually expanded in terms of the Gegenbauer polynomials
$C^{3/2}_n$ and $C^{1/2}_n$, respectively:
\bea\label{Ge1}
\phi^{\cal A}_{2;M} &=&
[\phi^{\cal A}_{2;M}]_{\rm as}(x) \biggl[ 1 + \sum^{\infty}_{n=1} a^{\cal A}_{n,M}C^{3/2}_n(2x-1) \biggr],
\nonumber\\
\phi^{\cal P}_{3;M} &=&
 [\phi^{\cal P}_{3;M}]_{\rm as}(x) \biggl[ 1 + \sum^{\infty}_{n=1} a^{\cal P}_{n,M}C^{1/2}_n(2x-1) \biggr],
\eea
where $[\phi^{\cal A}_{2;M}]_{\rm as}(x)=6x(1-x)$ and $[\phi^{\cal P}_{3;M}]_{\rm as}(x)=1$. The coefficients
$a^{\cal A(P)}_{n,M}$ are called the Gegenbauer moments and can be obtained from~\cite{NK06}
\bea\label{Gel2}
a^{\cal A}_{n,M}(x) &=& \frac{4n+6}{3n^2+9n+6}\int^1_0 dx\; C^{3/2}_n(2x-1)\phi^{\cal A}_{2;M}(x),
\nonumber\\
a^{\cal P}_{n,M}(x) &=& (2n+1) \int^1_0 dx\; C^{1/2}_n(2x-1)\phi^{\cal P}_{3;M}(x).
\eea
The Gegenbauer moments with $n>0$ describe how much the DAs deviate from the asymptotic one.
In addition to the Gegenbauer moments, we can also define the expectation value of the longitudinal
momentum, so-called $\xi(=2x-1)$-moments, as follows:
\be\label{Ge3}
\la\xi^n\ra^{\cal A(P)}_M = \int^1_0 dx\; \xi^n \phi^{\cal A(P)}_{2(3);M}(x).
\ee

\begin{table}[t]
\caption{The Gegenbauer moments and $\xi$ moments of twist-2 and twist-3 pion DAs obtained from the
linear and HO potential models compared other model estimates. }
\label{t2}
\renewcommand{\tabcolsep}{1pc} 
\begin{tabular}{@{}cccccccc} \hline\hline
Models &  Twists &  $a^{\cal A(P)}_{2,\pi}$ & $a^{\cal A(P)}_{4,\pi}$ & $a^{\cal A(P)}_{6,\pi}$
& $\la\xi^2\ra^{\cal A(P)}_\pi$ & $\la\xi^4\ra^{\cal A(P)}_\pi$ & $\la\xi^6\ra^{\cal A(P)}_\pi$ \\
\hline
HO & $\phi^{\cal A}_{2;\pi}$ & 0.0514 &  -0.0340 &  -0.0261 &  0.2176 &  0.0939 &  0.0508\\
     & $\phi^{\cal P}_{3;\pi}$ & -0.5816 &  -0.4110 &  -0.1725 & 0.2558 &  0.1231 & 0.0723\\
Linear & $\phi^{\cal A}_{2;\pi}$ & 0.1234 & -0.0033 &  -0.0218 & 0.2423 &  0.1136 & 0.0658\\
     & $\phi^{\cal P}_{3;\pi}$ & -0.3979 &  -0.3739 &  -0.2500 & 0.2803 &  0.1450 & 0.0907 \\
SR~\cite{Ball99} & $\phi^{\cal P}_{3;\pi}$ & 0.5158 &  0.2545 &  0.2162 & $\cdots$ &  $\cdots$ & $\cdots$ \\
SR~\cite{BBL} & $\phi^{\cal P}_{3;\pi}$ & 0.4373 &  -0.0715 &  -0.1969 & 0.3865 &  0.2451 & 0.1788 \\
SR~\cite{HWZ04} & $\phi^{\cal P}_{3;\pi}$ & $\cdots$ &  $\cdots$ &  $\cdots$
& $0.340\sim 0.359$ &  $0.164\sim 0211$ & $\cdots$ \\
SR~\cite{HZW05} & $\phi^{\cal P}_{3;\pi}$ & $\cdots$ &  $\cdots$ &  $\cdots$
& $0.52\pm 0.03$ &  $0.44\pm 0.01$ & $\cdots$ \\
$\chi$QM~\cite{NK06} & $\phi^{\cal P}_{3;\pi}$ & -0.4307 &  -0.5559 &  -0.1784 & 0.2759 &  0.1367 & 0.0816 \\
\hline\hline
\end{tabular}
\end{table}
In Table~\ref{t2}, we list the calculated Gegenbauer moments and $\xi$ moments of twist-2 and twist-3 pion DAs obtained from the linear and HO potential models at the aforementioned scale $\mu\sim 1$ GeV. Although the results of twist-2 pion DA were listed in our previous
work~\cite{CJ_DA}, we list them here again by
increasing the significant figures for completeness of this work. We also compare our results of
twist-3 DAs with other model estimates calculated at the scale $\mu=1$ GeV, e.g.
QCD sum-rules~\cite{Ball99,BBL,HWZ04,HZW05} and the chiral quark model ($\chi$QM)~\cite{NK06}.
As expected from the isospin symmetry, all the odd Gegenbauer and $\xi$ moments vanish.
It is interesting to note within our LFQM predictions that $a^{\cal P}_{2,\pi}$ of the twist-3 DA are negative
while the second Gegenbauer moments $a^{\cal A}_{2,\pi}$ of the twist-2 DA are positive,
regardless of the linear or the HO model parameters. Compared to other models for the twist-3
case, our results are quite different from those of QCD sum-rules~\cite{Ball99,BBL,HWZ04,HZW05}
but consistent with the chiral quark model predictions~\cite{NK06}. Again, the differences
between our LFQM and QCD sum-rule may be attributed to different treatment of constituent quark masses
as we discussed about the results shown in Fig.~\ref{fig4}.

\begin{table}[t]
\caption{The Gegenbauer moments and $\xi$ moments of twist-2 and twist-3 $K$ meson DAs obtained from the
linear and HO potential models compared other model estimates. }
\label{t3}
\renewcommand{\tabcolsep}{1pc} 
\begin{tabular}{@{}cccccccc} \hline\hline
Models &  Twists &  $a^{\cal A(P)}_{1,K}$ & $a^{\cal A(P)}_{2,K}$ & $a^{\cal A(P)}_{3,K}$
& $a^{\cal A(P)}_{4,K}$ & $a^{\cal A(P)}_{5,K}$ & $a^{\cal A(P)}_{6,K}$ \\
\hline
HO & $\phi^{\cal A}_{2;K}$ & 0.1316 &  -0.0278 &  0.0381 &  -0.0335 &  -0.0112 &  -0.0122\\
     & $\phi^{\cal P}_{3;K}$ & 0.3187 &  -0.7800 &  -0.0647 & -0.2923 & -0.2223 & -0.0396\\
Linear & $\phi^{\cal A}_{2;K}$ & 0.0894 & 0.0275 &  0.0575 & -0.0243 &  0.0069 & -0.0142\\
     & $\phi^{\cal P}_{3;K}$ & 0.2662 &  -0.6104 &  0.0486 & -0.3361 &  -0.1454 & -0.1161 \\
SR~\cite{Ball99}& $\phi^{\cal P}_{3;K}$ & $\cdots$ &  0.2631 &  $\cdots$ & -0.0522 &  $\cdots$ & 0.1470\\
SR~\cite{BBL}& $\phi^{\cal P}_{3;K}$ & 0.1837 &  0.2707 &  0.3953 & -0.2469 &  0.0550 & -0.2436\\
$\chi$QM~\cite{NK06} & $\phi^{\cal P}_{3;K}$ & 0.0236 &  -0.6468 &  -0.0367 & -0.3724 &  -0.0200 & -0.0940 \\
\hline\hline
 Models  & Twists  &  $\la\xi^1\ra^{\cal A(P)}_K$ & $\la\xi^2\ra^{\cal A(P)}_K$
 & $\la\xi^3\ra^{\cal A(P)}_K$
& $\la\xi^4\ra^{\cal A(P)}_K$ & $\la\xi^5\ra^{\cal A(P)}_K$ & $\la\xi^6\ra^{\cal A(P)}_K$ \\
\hline
HO & $\phi^{\cal A}_{2;K}$ & 0.0790 &  0.1905 &  0.0411 &  0.0759 &  0.0248 &  0.0389\\
     & $\phi^{\cal P}_{3;K}$ & 0.1062 &  0.2293 &  0.0600 & 0.1034 & 0.0389 & 0.0582\\
Linear & $\phi^{\cal A}_{2;K}$ & 0.0536 & 0.2094 &  0.0339 & 0.0895 &  0.0231 & 0.0486\\
     & $\phi^{\cal P}_{3;K}$ & 0.0887 &  0.2519 &  0.0560 & 0.1217 &  0.0394 & 0.0725 \\
SR~\cite{BBL} & $\phi^{\cal P}_{3;K}$ & 0.0612 &  0.3676 &  0.0593 &  0.2236 &  0.0520 & $\cdots$  \\
SR~\cite{HZW05} & $\phi^{\cal P}_{3;K}$ & $-0.10\pm 0.03$ & $0.43\pm 0.04$ &  $\cdots$ &  $\cdots$ &  $\cdots$ & $\cdots$  \\
$\chi$QM~\cite{NK06} & $\phi^{\cal P}_{3;K}$ & 0.0079 &  0.2471 &  0.0026 & 0.1166 &  0.0008 & $\cdots$ \\
\hline\hline
\end{tabular}
\end{table}
In Table~\ref{t3}, we display the calculated Gegenbauer moments and $\xi$ moments of twist-2 and twist-3 $K$ meson DAs obtained from the linear and HO potential models and compare them with other model predictions.
For the kaon case, the odd moments are nonzero due to the flavor SU(3) symmetry
breaking and the first moments $a^{\cal A(P)}_{1,K}$ is proportional to the difference between
the longitudinal momenta of the strange and nonstrange quarks in the two-particle Fock component.
We note within our LFQM predictions that the SU(3) symmetry breaking effects are rather significant
for the twist-3 DA than for the twist-2 DA~\cite{CJ_DA}.
Our results for the twist-3 $\phi^{\cal P}_{3;K}$ are overall in good agreement with those of the $\chi$QM~\cite{NK06}
except the values of the first moment $a^{\cal P}_{1,K}$ and $\la\xi^1\ra^{\cal P}_K$
with an order-of-magnitude difference between the two models, which may be understandable because the degree
of SU(3) symmetry breaking in $\chi$QM~\cite{NK06} is rather small compare to our LFQM
prediction. The shape of $\phi^{\cal P}_{3;K}$ obtained from $\chi$QM~\cite{NK06} is very
close to the symmetric and flat shape while the corresponding result from our LFQM has a rather sizable asymmetric form.

\begin{figure}
\begin{center}
\includegraphics[height=10cm, width=10cm]{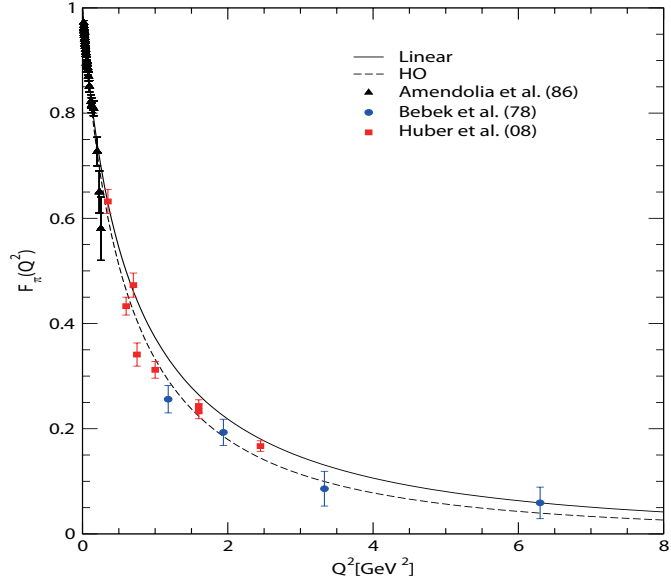}
\caption{\label{fig5} Pion electromagnetic form factors obtained from
$F^{\rm SLF (+)}_\pi =[F_\pi]^{\rm SLF(\perp)}_{\rm on}$ using the linear (solid line)
and HO (dashed line)  potential models.}
\end{center}
\end{figure}
In Fig.~\ref{fig5}, we show our numerical results of the pion electromagnetic form factor
from $F^{\rm SLF (+)}_\pi =[F_\pi]^{\rm SLF(\perp)}_{\rm on}$ using the linear (solid line)
and HO (dashed line)  potential parameters and compare with the available experimental
data~\cite{Am86,Be78,Hu08} up to the $Q^2\sim 8$ GeV$^2$ region.
In our previous LFQM analysis of the pion form factor~\cite{CJ06,CJ08}, we have also
shown that the usual power-law behavior of the pion form factor obtained in the perturbative
QCD analysis can also be attained by taking negligible quark masses in our nonperturbative LFQM,
confirming the anti-de Sitter space geometry/conformal field theory (AdS/CFT)
correspondence~\cite{BT04}.

\section{Summary and Discussion}
\label{sec:V}

As the zero-mode contribution is locked into a single point of the LF longitudinal momentum in the meson decay process,
one of the constituents of the meson carries the entire momentum of the meson and it is important to capture the effect from
a pair creation of particles with zero LF longitudinal momenta from the strongly interacting vacuum.
The LFQM with effective degrees of freedom represented by the constituent quark and antiquark
may thus provide the view of effective zero-mode cloud around the quark and antiquark inside the meson.
Consequently, the constituents dressed by the zero-mode cloud may be expected to satisfy the chiral symmetry
of QCD. Our results of this work for pseudoscalar mesons and the previous work for vector mesons were consistent with this expectation and effectively indicated that the constituent quark and antiquark in the standard LFQM could be considered as the dressed constituents including the zero-mode quantum fluctuations from the QCD vacuum.

In particular, we have discussed a
wave function dependence of the LF zero-mode contributions to the twist-3 two-particle
DA $\phi^{\cal P}_{3;M}$ of a pseudoscalar meson
between the two models, i.e. the exactly solvable manifestly
covariant BS model and the more phenomenologically accessible realistic LFQM using the
the standard LF (SLF) approach following our previous work~\cite{CJ_V14}.
As the SLF approach within the LFQM by itself is not amenable to determine the zero-mode contribution, we utilized the covariant BS model to check the existence (or absence) of the zero-mode.
Performing a LF calculation in the covariant BS model using the multipole type $q{\bar q}$ bound
state vertex function, we found that the twist-3 $\phi^{\cal P}_{3;M}$ receives
both the zero-mode and the instantaneous contributions and identified the zero-mode operator corresponding
to the zero-mode contribution.
We then linked the covariant BS model to the standard LFQM following the same correspondence
relation Eq.~(\ref{Eeq:12}) between the two models that we found in the vector meson decay
amplitude~\cite{CJ_V14} and
substituted the LF vertex function in the covariant BS model
with the more phenomenologically accessible Gaussian wave function provided by the LFQM analysis of meson mass~\cite{CJ_99}.
The remarkable finding is that the zero-mode contribution as well as the instantaneous
contribution revealed in the covariant BS model become absent in the LFQM with the Gaussian wave function.
Without engaging any of those treacherous contributions, our LFQM result of twist-3 DA $\phi^{\cal P}_{3;M}$
not only satisfies the fundamental constraint (i.e. symmetric form with respect to $x$) anticipated from the isospin symmetry
but also provides the consistency with the chiral symmetry (e.g. the correct
asymptotic form in the chiral symmetry limit) expected from the QCD.
This observation commensurates our previous observation made in the analysis
of vector meson decay process~\cite{CJ_V14}.

We have also shown that our treatment of the treacherous points
in two-point function is directly applicable to the three-point function,
analyzing the pion elastic form factor $F_\pi(Q^2)$ in the $q^+=0$ frame
both with the plus component ($J^+_{\rm em}$) and the perpendicular component ($J^\perp_{\rm em}$) of the current.
This analysis portrayed that the instantaneous contribution appeared in the covariant BS model became absent in
the LFQM. It supports the conclusion drawn from the analysis of the two-point function.

From the self-consistent covariant description of the twist-3 $\phi^{\cal P}_{3;M}$ together with the
previously obtained~\cite{CJ_DA} twist-2 DA $\phi^{\cal A}_{2;M}$ of a pseudoscalar meson in our LFQM, we presented
a good deal of numerical results obtained from our LFQM.
The quark condensate obtained from the normalization condition of $\phi^{\cal P}_{3;\pi}$, i.e.
$\la {\bar q}q\ra =-(285.8 {\rm MeV})^3 [-(263.7 {\rm MeV})^3]$ for the linear~[HO] potential parameters
comes out reasonable compared to the commonly used phenomenological value $-(250 {\rm MeV})^3$.
The ratio of the second transverse moments for the axial-vector and the pseudoscalar channels,
$\la {\bf k}^2_\perp \ra^{\cal A}_\pi / \la {\bf k}^2_\perp \ra^{\cal P}_\pi = 0.558~[0.597]$
 for the linear~[HO] parameters, is in good agreement with the QCD sum-rule result, 5/9~\cite{SR_C1}. Of particular interest, the mixed quark-gluon condensate of dimension 5
estimated from the value of $\la {\bf k}^2_\perp\ra^{\cal P}_\pi$~[see Eq.~(\ref{N3})],
$\la ig\bar{q}\sigma\cdot G q\ra=-(491.1~{\rm MeV})^5~[-(442.2~{\rm MeV})^5]$ for the
linear~[HO] parameters, is also in good
agreement with the result $-(490~{\rm MeV})^5$ of the direct calculation in
the instanton model~\cite{PW}.
Moreover, our numerical results of $\phi^{\cal P}_{3;\pi}$ not only show the symmetric forms
anticipated from the isospin symmetry but also reproduce the exact asymptotic result
$[\phi^{\cal P}_{3;\pi}]_{\rm as}(x)=1$ in the chiral symmetry ($m_q\to 0$) limit.
For the kaon case, the results of $\phi^{\cal P}_{3;K}$ show asymmetric form as expected from the
flavor SU(3) symmetry breaking. Our results for the
Gegenbauer moments and $\xi$ moments of
twist-3 pion and kaon DAs are overall in good agreement with the chiral quark model~\cite{NK06}
although they differ from those of QCD sum-rule estimates~\cite{Ball99,BBL,HZW05}.

For further analysis, it would be interesting to study this process with other vertex function such as
the symmetric product ansatz suggested in Eq. (38) of Ref.~\cite{FPPS}.
The generalization of our findings to the three-point function would also require
the analysis of unequal quark and antiquark mass cases.

\acknowledgments
This work was supported by the Korean Research Foundation
Grant funded by the Korean Government (KRF-2010-0009019).
C.-R. Ji was supported in part by the US Department of Energy
(Grant No. DE-FG02-03ER41260).


\begin{thebibliography}{99}
\bibitem{BL80} G.P. Lepage and S.J. Brodsky, \Journal{\PRD}{22}{2157}{1980}.

\bibitem{ER80} A.V. Efremov and A.V. Radyushkin, \Journal{\PLB}{94}{245}{1980}.

\bibitem{CZ84} V.L. Chernyak and  A.R. Zhitnitsky, Phys. Rep. {\bf 112}, 173 (1984).

\bibitem{BF} V. M. Braun and I.E. Filyanov, Z. Phys. C {\bf 48}, 239 (1990).

\bibitem{Ball99} P. Ball,  J. High Energy Phys. 01, 010 (1999).

\bibitem{BBL} P. Ball, V. M. Braun, and A. Lenz, J. High Energy Phys. 05, 004 (2006).

\bibitem{CJ_DA} H.-M. Choi and C.-R. Ji,
\Journal{\PRD}{75}{034019}{2007}.

\bibitem{PR} M. Praszalowicz and A. Rostworowski, Phys. Rev. D {\bf 66}, 054002 (2002).

\bibitem{NK06} S.I. Nam and H.-Ch. Kim, \Journal{\PRD}{74}{096007}{2006}.

\bibitem{HWZ04} T. Huang, X.H. Wu, and M.Z. Zhou, \Journal{\PRD}{70}{014013}{2004}.

\bibitem{HZW05} T. Huang, M.Z. Zhou, and X.H. Wu, \Journal{\EPJC}{42}{271}{2005}.

\bibitem{BPP} S. J. Brodsky, H. -C. Pauli, and S. Pinsky, Phys. Rep. {\bf 301}, 299 (1998).


\bibitem{Zero1} S. J. Brodsky and D. S. Hwang, \Journal{\NPB}{543}{239}{1999}.

\bibitem{Zero2} J.P.B.C. de Melo, J.H.O. Sales, T. Frederico, and P.U. Sauer, \Journal{\NPA}{631}{574c}{1998}.

\bibitem{Zero3} H.-M. Choi and C.-R. Ji, \Journal{\PRD}{58}{071901(R)}{1998}.

\bibitem{CJ_99} H.-M. Choi and C.-R. Ji,
\Journal{\PRD}{59}{074015}{1999}; \Journal{\PLB}{460}{461}{1999}.

\bibitem{CJ_Bc} H.-M. Choi and C.-R. Ji, \Journal{\PRD}{80}{054016}{2009}.

\bibitem{Jaus99} W. Jaus, \Journal{\PRD}{60}{054026}{1999}.

\bibitem{Jaus03} W. Jaus, \Journal{\PRD}{67}{094010}{2003}.

\bibitem{Cheng04} H.-Y. Cheng, C.-K. Chua, and C.-W. Hwang,
\Journal{\PRD}{69}{074025}{2004}.

\bibitem{CJ_PV} H.-M. Choi and C.-R. Ji,
\Journal{\NPA}{856}{95}{2011}; \Journal{\PLB}{696}{518}{2011}.

\bibitem{MF12} J.P.B.C. de Melo and T. Frederico, \Journal{\PLB}{708}{87}{2012}.

\bibitem{Jaus90} W. Jaus, \Journal{\PRD}{41}{3394}{1990}.

\bibitem{Choi07} H.-M. Choi, \Journal{\PRD}{75}{073016}{2007}.

\bibitem{CJ_V14} H.-M. Choi and C.-R. Ji, \Journal{\PRD}{89}{033011}{2014}.

\bibitem{BDJM} B.L.G. Bakker, M.A. DeWitt, C-R. Ji and Y. Mishchenko,
Phys. Rev. D {\bf 72}, 076005 (2005).

\bibitem{JMT2013} C.-R. Ji, W. Melnitchouk and A.W. Thomas,
Phys. Rev. Lett. {\bf 110}, 179101 (2013).

\bibitem{Kon} L. A. Kondratyuk and D. V. Tchekin, Physics of Atomic Nuclei {\bf 64}, 727 (2001).

\bibitem{Cheng97} H.-Y. Cheng, C.-Y. Cheung, and C.-W. Hwang,
\Journal{\PRD}{55}{1559}{1997}.

\bibitem{Hwang10} C.-W. Hwang, \Journal{\PRD}{81}{114024}{2010}.

\bibitem{Jaus91} W. Jaus, \Journal{\PRD}{44}{2851}{1991}.

\bibitem{CCP} P. L. Chung, F. Coester, and W. N. Polyzou,
\Journal{\PLB}{205}{545}{1988}.

\bibitem{Card95} F. Cardarelli, I.L. Grach, I.M. Narodetskii, G. Salme,
S. Simula, \Journal{\PLB}{349}{393}{1995}.

\bibitem{CJK02} H.-M. Choi, C.-R. Ji, and L.S. Kisslinger,
\Journal{\PRD}{65}{074032}{2002}.

\bibitem{GOR} M. Gell-Mann, R. Oakes, and B. Renner, Phys. Rev. {\bf 175}, 2195 (1968).

\bibitem{BCJ01} B.L.G. Bakker, H.-M. Choi, and C.-R. Ji,
 \Journal{\PRD}{63}{074014}{2001}.

\bibitem{CJK}  L.S. Kisslinger, H.-M. Choi, and C.-R. Ji,
\Journal{\PRD}{63}{113005}{2001}.

\bibitem{CJ06} H.-M. Choi and C.-R. Ji,
\Journal{\PRD}{74}{093010}{2006}.

\bibitem{CJ08} H.-M. Choi and C.-R. Ji,
\Journal{\PRD}{77}{113004}{2008}.

\bibitem{BT04} S.J. Brodsky and Guy F. de Teramond, \Journal{\PLB}{582}{211}{2004};
\Journal{\PRL}{96}{201601}{2006}.

\bibitem{MF97} J.P.B.C. de Melo and T. Frederico, \Journal{\PRC}{55}{2043}{1997}.

\bibitem{SS} C.M. Shakin and W.-D. Sun, \Journal{\PRC}{51}{2171}{1995}.

\bibitem{MFPS02}
J.P.B.C. de Melo, T. Frederico, E.Pace, and G. Salm\'{e},
\Journal{\NPA}{707}{399}{2002}.

\bibitem{BCJ03} B.L.G. Bakker, H.-M. Choi, and C.-R. Ji,
 \Journal{\PRD}{67}{113007}{2003}.

\bibitem{Melosh} H. J. Melosh, \Journal{\PRD}{9}{1095}{1974}; P. L. Chung, F. Coester, B. D. Keister, and
W. N. Polyzou, \Journal{\PRC}{37}{2000}{1988}.

\bibitem{PDG}  K.A. Olive {\em et al}. (Particle Data Group),
 Chin. Phys. C {\bf 38}, 090001 (2014).

\bibitem{SR_C1} V.L. Chernyak, A.R. Zhitnitsky, and I.R. Zhitnitsky, Yad. Fiz. {\bf 38}, 1074 (1983)
[Sov. J. Nucl. Phys. {\bf 38}, 645 (1983)].

\bibitem{SR_C2} A.R. Zhitnitsky, Phys. Lett. B {\bf 329}, 493 (1994).

\bibitem{PW} M.V. Polyakov and C. Weiss, Phys. Lett. B {\bf 387}, 841 (1996).

\bibitem{Am86} S. R. Amendolia {\em et al.}, \Journal{\NPB}{277}{168}{1986}.

\bibitem{Be78} C. J. Bebek {\em et al.}, \Journal{\PRD}{17}{1693}{1978}.

\bibitem{Hu08} C. M. Huber {\em et al.}, \Journal{\PRC}{78}{045203}{2008}.

\bibitem{FPPS} T. Frederico, E. Pace, B. Pasquini, and G. Salme,
 \Journal{\PRD}{80}{054021}{2009}.

\end{thebibliography}
\end{document}